\def\VEL{\:{\rm km\:s^{-1}}}

\def\OIGS{\:{\rm ergs\:cm^{-2}\:s^{-1}\:\AA^{-1}}}

\def\SviL{\ion{S}{6} $\lambda\lambda933,945$}

\documentclass[12pt,preprint]{aastex}

\begin{document}

% Additional private definitions that appear to work only inside document

\newcommand{\MSOL}{\mbox{$\:M_{\sun}$}}

\newcommand{\EXPN}[2]{\mbox{$#1\times 10^{#2}$}}
\newcommand{\EXPU}[3]{\mbox{\rm $#1 \times 10^{#2} \rm\:#3$}}  % exponent with units
\newcommand{\POW}[2]{\mbox{$\rm10^{#1}\rm\:#2$}}

% End of defining things

\title{\bf{WZ Sge: {\it FUSE} Spectroscopy of the 2001
Outburst\footnote{Based on observations made with the
NASA-CNES-CSA Far Ultraviolet Spectroscopic Explorer. {\it FUSE}
is operated for NASA
by the Johns Hopkins University under NASA contract NAS5-32985}
%\\ \today
}
}

\author{Knox S. Long \& Cynthia S. Froning}

\affil{Space Telescope Science Institute, \\ 3700 San Martin
Drive, \\ Baltimore, MD 21218}

\author{Boris G\"ansicke \& Christian Knigge}

\affil{Department of Physics \& Astronomy,\\
University of Southampton, \\
Southampton SO17 1BJ UK}

\author{Edward M. Sion}

\affil{Department of Astronomy \& Astrophysics,\\
Villanova University, \\
Villanova, PA, 19085}

\and

\author{Paula Szkody}

\affil{Department of Astronomy,\\
University of Washington, \\
Seattle, WA 98195}

\begin{abstract}
The short-period cataclysmic variable WZ Sge underwent its first
dwarf nova outburst in 23 years in 2001 July.  Here we describe
905-1187 \AA\ spectra obtained with the Far Ultraviolet
Spectroscopic Explorer ({\it FUSE}) during the outburst and within
the first few months after the outburst. The first spectrum,
obtained about 7 days after the outburst began and 6.4 days past
optical maximum, is dominated by a broad (50-60 \AA\ EW) feature
due to O VI and can be characterized in terms of emission from a
highly inclined disk modulated by the effects of a highly ionized
wind. WZ Sge is fainter in later observations (starting about 45
days after optical maximum) and increasingly ``red'' in the far
ultraviolet. The spectral shape increasingly resembles that
expected from a metal-enriched white dwarf photosphere, although
it seems likely that some disk contribution remains at least
through the second {\it FUSE} observation, when the visible light
curve showed the system to be undergoing a rebrightening event.
The absorption lines observed on 29 September and in early
November, when the visible light curve was in a fairly steady
decline toward quiescence, are narrow, with widths (FWHM) of order
0.7-1.3 \AA\ (200-400 $\VEL$). The most natural explanation for
these spectra is that the continuum arises from a relatively
massive, slowly rotating white dwarf, and that the lines are
created by a combination of metals in the atmosphere and of
absorbing material along the line of sight to the white dwarf.  If
the white dwarf has $\log{g}= 8.5$, then the white dwarf cooled
from $\sim$25,000 K on 29 September to 23,000 K in early November.

\end{abstract}


\section{Introduction}
Cataclysmic variables (CVs) are mass-exchanging binary star
systems containing a white dwarf (WD) and a relatively normal
main-sequence companion. Dwarf novae are a class of CVs that
exhibit outbursts lasting from a day to a few weeks, depending on
the system. Usually, outburst amplitudes are 3-5 magnitudes above
quiescence and times between outbursts range from a few weeks to a
few months. The outbursts arise from a large increase in the
effective temperature and optical depth of the disk that mediates
the mass transfer from the normal star to the WD. All CVs,
including dwarf novae, are thought to have evolved from more
widely separated binaries that have passed through a common
envelope phase, and are thought, as a result of angular momentum
loss, to evolve toward periods of about 80 minutes on Gyr
timescales \citep{rappaport1982}.
%The contraction of the systems
%is not driven directly by the accretion. Early on, a phenomenon
%known as magnetic braking is likely to be involved, but in shorter
%period systems gravitational radiation is thought to be the
%driving mechanism. However, gravitational radiation ceases to be
%effective in bringing the system together at the period when the
%Kelvin-Helmholtz timescale of the secondary exceeds the
%gravitational radiation time scale, and in fact the period begins
%to lengthen.
Population synthesis calculations suggest that most of the CVs in
the Galaxy should be at or near the period minimum. But the mass
transfer rate of systems near the minimum is low. Systems near the
period minimum are therefore expected to be faint in quiescence,
to have larger than normal outburst amplitudes, and to have
exceptionally long intervals between outbursts. As a result,
detected systems near the period minimum are rare.

With a parallax-determined distance of 44 (+6.5,-5) pc
\citep{thorstensen2001} and an orbital period of 82 minutes, WZ
Sge is the closest and best studied example of a CV near the
period minimum. The system is view at a relatively high
inclination angle \citep[$\sim75\degr$,][]{smak1993}; the
secondary eclipses the bright spot and the outer portion of the
disk, but not the WD. Prior to July of 2001, it had been observed
to go into outburst only 3 times, once in 1913, once in 1946, and
most recently in 1978. During the last outburst it was observed
extensively with {\it IUE}
\citep{fabian1980,friedjung1981,naylor1989}. The spectra at the
peak of the outburst resembled those of other outbursting systems;
the spectral slope was consistent with that expected from a
steady-state accretion disk in outburst. After outburst, the
system was not observed for about 180 days, but when observations
recommenced, the spectrum had evolved into something that
resembled a WD cooling from 20,500 K to 15,400 K with an e-folding
timescale of about 690 days \citep{slevinsky1999}. WZ Sge was
observed with {\it HST} in the early-mid 1990s and
\cite{cheng1997} estimated a temperature of 14,800 K. They also
found that the WD was rotating with $v\:sin(i)$ of 1200
(+300/-400) $\VEL$ and that rapid rotation was coupled with
abundances (most notably supersolar C) unlike those observed in
any other WD in a dwarf nova system. In addition, \cite{welsh1997}
found pulsations in the UV lightcurve in high time resolution FOS
data with a period of 28 s, in the range where periodicities have
also been observed at optical and X-ray wavelengths
\citep{patterson1998}.\footnote{The {\it HST} observations of the
the 2001 outburst have revealed large (5\%) amplitude oscillations
at UV wavelengths within a month of outburst maximum
\citep{knigge2002}. This was the first time that periodicities
near 15 s have been seen to be the dominant oscillation.} The fact
that the oscillations are not always present and are unstable in
period has led to a lively debate as to whether WZ Sge is
\citep{patterson1998} or is not \citep{skidmore1999} a DQ Her
star, a class of CVs whose magnetic field channels material from
the accretion disk onto selected regions of the WD surface.

The CV WZ Sge began a dwarf nova outburst on 23 July, 2001, about
10 years earlier than predicted based on the frequency of the
previously known outbursts of the system.  Extensive ground and
space-based efforts were mounted in an attempt to better
understand WZ Sge. These included a series of observations with
the {\it Far Ultraviolet Spectroscopic Explorer} ({\it FUSE}),
which provided high resolution (R$\sim$12000) spectra in the
wavelength range 905-1187 \AA. Here we present an initial
description of those observations, and an analysis of the
emergence of the WD in the post-outburst spectra. A detailed
discussion of the outburst spectrum will be reported elsewhere.

\section{Observations and qualitative description of the data}

The {\it FUSE} satellite was launched into low-earth orbit in June
1999 \citep{moos2000}. The {\it FUSE} optical system consists of
four optical telescopes that feed four Rowland spectrographs
producing 4 independent spectra on two photon counting area
detectors.  The optics for two of the four channels are coated
with Si:C optimized for shorter (905-1105 \AA) wavelengths and two
with Al:LiF for longer ($\sim$1000-1187 \AA) wavelengths.  The
observatory, its operations, and performance have been described
in detail by \cite{sahnow2000}.

Following the announcement that an outburst of WZ Sge was underway
and the approval of  Director's discretionary time to observe WZ
Sge, the {\it FUSE} operations staff carried out four observations
of WZ Sge. VSNET and AAVSO lightcurves of the outburst indicate
that outburst maximum occurred near  0 hr UT 24 July (JD
2452114.5). As shown in Fig.\ \ref{wzsge_obs}, the optical
outburst can be characterized by a main outburst lasting 24 days
until 16 August (JD 2452138), a series of decline and
rebrightening phases lasting until mid-September, and then a
rather steady decline toward quiescence. As was typical of earlier
outbursts, WZ Sge reached a peak apparent magnitude of 8.2, more
than 7 magnitudes above its quiescent brightness of 15.5
\citep[see, e.g.][for a detailed discussion of the optical
lightcurve]{patterson2002}. The first {\it FUSE} observations
occurred about 6.4 days after outburst maximum. The second
observation occurred during the decline from one of the ``echo''
outbursts, while the last two appear to be part of the steady
decline.  The November observation, which was spaced out over five
days, was the last observation that could be scheduled until late
in 2002 due to spacecraft design and operational constraints. A
journal of the observations that occurred is shown in Table
\ref{tab_obs}. Total exposure times varied from 3,180 in the first
observation to 15,183 s in the last observation, roughly matching
the decline in flux during this period.

All of the observations were obtained using the LWRS (30\arcsec x
30\arcsec) aperture, which is least prone to slit losses due to
misalignments of the four {\it FUSE} telescopes. Because of
limitations in onboard data storage, the first two observations
were obtained in histogram mode, in which spectral images are
accumulated over a period of time and transmitted to the ground. A
total of eight spectra with individual exposure times ranging from
392 to 403 seconds were obtained on 30 July. For the 7 September
observations, there were 16 spectra, and exposure times ranged
from  460 to 593 seconds. However, the 7 September observation was
marred by the temporary HV shutdown of the detector 1 due to a
single bit upset event after the first four spectra were obtained,
thereby eliminating half of the spectral channels, and
complicating somewhat the analysis of the second observation as
described below.  Predicted count rates for WZ Sge were  low
enough in the 29 September and November observations to permit use
of the high time resolution mode, in which each event time and
position on the detector is recorded. There were no anomalies in
either of these observations.

An overview of the time-averaged spectra obtained in each of the
observations is shown in Fig.\ \ref{data_fourpanel}.  All of the
data shown here were reprocessed using V.2.0.5 of the {\it FUSE}
calibration pipeline; for faint sources, the flux calibration in
this version of the pipeline is significantly better than that
used to produce the calibrated spectra in the archive at the time
of the observations. As noted above, {\it FUSE} has four
spectroscopic apertures. During our observations, Fine Error
Sensor A, which images the LiF~1 aperture, was used to guide the
telescope. It is possible due to thermal drifts between the
channels for the source to drift partly or completely out of
non-guiding channels. Fortunately, the spectral regions covered by
the spectral channels overlap and therefore one can compensate for
this problem. Basically, we eliminated spectra from completely
misaligned channels and used the overlap regions to renormalize
the remaining spectra in the SiC~1, LiF~2 and SiC~2 to the flux in
the LiF~1 channel. To produce a final spectrum covering the full
{\it FUSE} wavelength range, we masked low sensitivity portions of
each channel (including the so-called ``worm" which complicates
LiF~1 longward of 1134 \AA), and combined the rescaled spectra.
When we combined, we weighted according to the area and exposure
time for that channel and then rebinned onto a common wavelength
scale with, in this case, 0.1 \AA\ resolution.

As is evident from Fig.\ \ref{data_fourpanel}, there was a major
evolution in both the FUV flux and the spectral shape of WZ Sge
during the course of the 4 observations. The flux at 1150 \AA\
declined from \EXPN{6}{-12} to \EXPN{9}{-13} to \EXPN{4}{-13} to
\EXPU{2}{-13}{\OIGS} on 30 July, 7 September, 29 September and
early November, respectively. The flux drop was greatest at the
shortest wavelengths; on 30 July the flux at the Lyman limit was
comparable to the flux at 1150 \AA\ but by November there was no
detectable emission at the Lyman limit.

WZ Sge was observed with {\it HST} on 8 August
\citep{kuulkers2002} and the flux observed with {\it HST} was
similar to that observed with {\it FUSE} on 30 July in the
wavelength ranges where the spectra overlap. The flux observed on
30 July with {\it FUSE} was somewhat larger than the
\EXPU{1-2}{-12}{\OIGS} reported by \cite{fabian1980} for the 1978
outburst using {\it IUE}, but the first {\it IUE} observation was
about 15 days after optical maximum.  The flux in November was
comparable to that reported by \cite{slevinsky1999} in the first
post-outburst observation from the 1978 outburst, even though that
observation occurred 222 days after outburst onset, while the {\it
FUSE} observation occurred about 105 days after outburst onset.
%The flux at 1400 \AA\ and at the short wavelength limit was about
%\EXPU{1.8}{-13}{\OIGS}. The first observation in quiescence took
%place on 10 July 1979.

\subsection{30 July -- WZ Sge in Outburst}

The first spectrum of WZ Sge is dominated by prominent O VI
emission.  The entire spectrum, as shown in Fig.\
\ref{wzsge1_lines}, is complex.  Many of the features can be
identified with transitions of C III, Si III, S III, S IV, or the
Lyman lines of hydrogen.  Most are absorption features, though
some, such as S VI, the Lyman lines of hydrogen, and C III
$\lambda$1178 have associated emission wings.

Although the exact placement of the continuum is difficult to
establish, as a result of the complexity of the spectrum, the
equivalent width of O VI is of order 50 \AA\ and the FWZI appears
to be at least 30 \AA, indicating a velocity spread of order 4000
$\VEL$. The Lyman$\gamma$ profile in the first spectrum shows
evidence of a red emission wing (see Fig.\ \ref{wzsge1_lines}) and
therefore it is possible that there is some admixture of
Lyman$\beta$ in what must be a predominately O VI profile.  The
emission feature is punctuated by several relatively narrow
absorption features, or self-reversals, with cores located at
1024.6, 1025.6, 1030.6, 1036.3, and 1039.2 \AA. The features at
1030.6, and 1036.3 are clearly due to O VI, blueshifted by 1.33
\AA\ or -380 $\VEL$. Since the $\gamma$ velocity if the system is
-72 $\VEL$ \citep{steeghs2001}, the narrow O VI features suggest
material flowing out of the system toward the observer. Simple
Gaussian fits to the two O VI absorption features indicate EW
(relative to the emission line) of about 1.9 \AA\ for both
features and FWHM of about 2 \AA; however both of the lines are
somewhat asymmetric in shape. The two features near 1025 \AA\ are
presumably both due to Lyman$\beta$, the longer feature being near
the rest wavelength of 1025.72 \AA\ for Lyman$\beta$, while the
other is blue-shifted by a similar amount to O VI. The feature is
not due to He II$\lambda$1025.27, which is separated by 0.54 \AA\
from Lyman$\beta$. A similar structure is seen in the
Lyman$\gamma$ profile. The feature at 1039.2 \AA\ is due to
interstellar O I.

If the continuum  flux level is about \EXPU{5}{-12}{\OIGS} in the
wavelength range 950-1020 \AA, then there is clear excess of
emission between 925 and 950 \AA. Similar excesses are observed in
UX UMa \citep{froning2002} and in the quiescent or near quiescent
spectra of SS Cyg \citep{long2002_sscyg}. The narrow absorption
features in the spectrum in this region are mainly due to
(interstellar) H I.  Many of the narrow features have associated
broader features, suggesting that H and (perhaps) He II emission
is heavily involved in the overall shape of the spectrum. The two
prominent peaks at 935 and 946 \AA\ are likely to be the P-Cygni
emission profiles of \SviL. The N IV $\lambda$923 multiplet, the
same multiplet that appears in C III at 1175 \AA\, is clearly
present in the outburst spectrum.  There is no emission shortward
of the Lyman limit.

There are variations in the {\it FUSE} outburst spectra on orbital
timescales. This is indicated in Fig.\ \ref{d1_lif1_O6}, which
shows each of the eight LiF~1 spectra of the O VI line region. The
overall flux in this region (and in the rest of the spectrum) is
least during the observation obtained at phase 0.04. To illustrate
the differences in the spectra we have faintly overlayed the
histogram obtained at phase 0.04 over all of the other spectra.
It is apparent from this comparison that the main variation is in
the continuum. The peak of the O VI line is relatively constant as
are flux levels in the centers of the line cores, especially those
of Si IV $\lambda\lambda 933,945$. Given that the largest fluxes
occur near orbital phase 0.83, it is tempting to associate the
changing flux with an orbital hump. This suggestion has to be
tempered however, with the realization that the number of spectra
is quite small, particularly near orbital phase 0, and therefore
one cannot rule out the possibility that we are seeing secular
variations. There are also some variations in the line profiles of
O VI; in particular in those periods in which the continuum flux
is high, there is a ``notch'' on the red side of the O VI
absorption profiles that is stronger in some spectra than in
others.

\subsection{7 September -- The spectrum in the rebrightening phase}

The character of the spectrum had changed considerably by the time
of the second observation.  As shown in Fig.\ \ref{wzsge2_lines},
the continuum is no longer flat (in $F_{\lambda}$), but turns down
at wavelengths shorter than 1050 \AA. O VI is no longer present in
emission; the broad absorption profile that is present near O VI
is centered on and dominated by Lyman$\beta$. O VI is still
present in absorption on the red wing of the broader Lyman$\beta$
profile. There are no indications of other emission features
associated with S VI, C III or the higher order Lyman lines
either.  The prominent lines in the spectrum have Gaussian FWHM of
order 1.3-1.7 \AA\ (350-500 $\VEL$). The lines are also
blue-shifted by 0.5-1 \AA\ (150-300 $\VEL$) compared to the
interstellar lines in the spectrum; this blue-shift is somewhat
greater than the systemic velocity of -72 $\VEL$
\cite{steeghs2001}.

Because of the shutdown of the detector 1 during the 7 September
observation, and hence the loss of the channel likely to provide
the best photometric accuracy, our ability to make judgements
about temporal variability are compromised. Nevertheless, it is
quite clear that the variations if they do exist are relatively
modest.  The standard deviation of the flux in the region around
1050 \AA\ is less than 15\%, with the largest deviation of order
30\%.  There are subtle variations in line shapes but they are
also quite modest.

\subsection{29 September and early November -- Spectra from the decline phase}

By the time of the 29 September observation, the last
rebrightening event was over and the apparent temperature of the
spectrum, as shown in Fig.\ \ref{wzsge3_lines} has dropped
further.  As was the case in early September, the Lyman lines are
quite evident, and the flux in the core of Lyman$\beta$ is now
very small.\footnote{There are some narrow emission features that
become more prominent in the fluxed spectra as WZ Sge fades.
However, these are all geocoronal and unassociated with WZ Sge.}
The region around O VI is fairly complex, and most likely involves
absorption and emission components, possibly arising from an
optically thin disk. With the exception of the Lyman lines, the
absorption lines are slightly narrower with FWHM of order 0.7-1.53
\AA\  (200-400 $\VEL$).

This trend toward lower apparent temperature continues in
November, as shown in Fig.\ \ref{wzsge4_lines}. On 29 September,
emission (at a flux level of order \EXPU{1}{-14}{\OIGS}) is
observed to the Lyman limit, but by November there is no
detectable emission at wavelengths shorter than 940 \AA.

Since the 29 September and November observations were obtained in
the TIME-TAG mode, we divided up the exposures and constructed
spectra in 200 s intervals, and searched for both secular trends
and changes associated with the orbital period. Neither set of
data shows large secular changes in the continuum flux, and
changes in absorption line strengths are modest.  In fact, the
only variability uncovered was on 3 November when the strength of
the absorption line spectrum appears to have increased during the
last 2400 seconds of that day's observation. This is shown in
Fig.\ \ref{data_var4}, which compares the spectrum in the LiF~1
channel, the channel used for guiding, during the period beginning
at 07:13 UT to the spectrum during the rest of the November
observation. The continuum appears to be unchanged, but the
absorption lines were clearly deeper in this interval than during
the rest of the November observations.

\section{Analysis and Discussion}

\subsection{30 July -- Outburst}

In outburst, the spectra of dwarf novae are generally understood
to be dominated by disk emission.  However, as has been clear
since early {\it IUE} observations,
\citep[e.g.][]{heap1978,cordova1982}, most dwarf novae in outburst
have spectral features (such as C IV) that reflect scattering of
disk photons in high velocity outflows emerging from the inner
disk. More recently detailed analyses of UV spectra of some CVs
\citep{knigge1997} and time-resolved spectroscopy of other dwarf
novae \citep{horne1994} have made it clear that dwarf novae have
additional material elevated at significant distance above the
disk plane. This material has been interpreted variously as a disk
chromosphere, as a thickening of the disk at its outer edges (or
the co-rotation radius), or as material from the secondary
overflowing the disk edge.  Because of the plethora of resonance
lines in the {\it FUSE} wavelength range, {\it FUSE} is quite
sensitive to the existence of this material.

The {\it FUSE} spectrum of WZ Sge in outburst needs to be
interpreted in this context. Qualitatively, the {\it FUSE}
spectrum is somewhat similar to that obtained of the novalike
variable UX UMa \citep{froning2002}, and quite different from the
two long-period dwarf novae in outburst that we have studied (U
Gem; Froning et al.\ 2001 and SS Cyg; Long et al.\ 2002). In
particular, the outburst spectra of WZ Sge in the {\it FUSE}
region are complex, with narrower absorption lines than expected
from a steady-state accretion disk viewed at high inclination.
However, to our knowledge, the prominence of O VI makes the WZ Sge
spectrum unique.  Both the O VI line and the other features are
likely to be due to a wind and/or corona vertically extended above
the disk in WZ Sge, and the strength of O VI may be due to
ionization of the wind by X-rays created in the hot boundary layer
\citep{kuulkers2002}. It is ultimately our intention to attempt to
model the outburst spectra of WZ Sge in detail in an attempt to
determine the kinematic and ionization structure of the line
producing regions, but this will require significant effort.

Here, we limit ourselves to an approximate estimate of the
accretion rate. Specifically, we have compared the observed
spectrum to model spectra created for steady-state disks. The
model spectra themselves were constructed from summed,
area-weighted, Doppler broadened stellar atmospheres, assuming the
standard steady-state temperature distribution in the disk
\cite[see, e.g.][]{froning2001}. For the parameters of the system,
we assumed a non-rotating 0.8 \MSOL WD \citep{skidmore2000}, a
disk inclined at 75\degr\ \citep{smak1993}, and a boundary layer
that did not contribute significantly to the continuum.

We have made estimates of the accretion rate both from the flux
observed, assuming a distance of 44 pc, and from the shape of the
spectrum. For the 30 July observation, the flux is consistent with
an overall accretion rate of \EXPU{8.5}{-10}{\MSOL yr^{-1}}, while
a fit to the spectral shape indicates a value roughly 3 times
larger.\footnote{These fits were made for an assumed WD mass and
radius appropriate for a non-rotating WD. Larger values of the
accretion rate would result if the WD was of lower mass, both
because the gravity is less and because the radius of a lower mass
WD is larger.}  While the model shown in Fig.\ \ref{wzsge1_disk}
is not a good fit to the data, it does have a slope which
plausibly could underly a more complicated spectrum. It is also
painfully obvious that standard disk models will not accommodate
the narrow features in the spectrum. This is because features in
the disk photosphere are broadened by the Doppler rotation of the
inner disk at the inclination appropriate for WZ Sge. Nonetheless,
we believe it is significant that the values of the disk accretion
rate $\dot{m}_{disk}$ are lower than is often estimated for dwarf
novae in outburst, but are consistent with its relatively modest
brightness compared to other dwarf novae at well-known distances.
For example, U Gem has an astrometric distance of $96.4\pm4.6$ pc
\citep{harrison1999} and an inclination angle of $67\pm3\degr$
\citep{long1999}, similar to WZ Sge. {\it FUSE} spectra obtained
of U Gem during the peak of an outburst in 2000 March show flux of
\EXPU{1.8}{-11}{\OIGS} at 1150 \AA, roughly 3 times that of WZ
Sge. \cite{froning2001} estimated the mass accretion rate to be
\EXPU{7}{-9}{\MSOL ~ yr^{-1}}, almost 10 times that of the
flux-based estimate for WZ Sge (and 3 times greater than the
spectral shape-based estimate).  This conclusion has to be
mitigated somewhat due the fact that WZ Sge had already declined
from maximum by about 1.1 magnitudes by 30 July. This {\it FUSE}
spectrum was the first obtained of the outburst, and one cannot
state with confidence that the flux was not significantly higher
earlier. On the other hand, there was very little evolution of the
UV flux between 30 July and 8 August, when the first {\it HST}
observation took place, and the optical magnitude had declined by
an additional 0.9 magnitudes.

\subsection{29 September and Early November -- Post-outburst spectra\label{wd}}

As discussed in section 1, the UV spectra of WZ Sge in quiescence
have been interpreted in terms of emission from a rapidly-rotating
WD that cooled from about 20,500 K about 180 days after outburst
to about 15,000 K 15 years later \citep{slevinsky1999,cheng1997}.
%This is consistent with our understanding that DN outbursts result
%from a transition of the disk (or disk photosphere) from a cold,
%unionized to hot ionized state, since in that case the disk should
%not be a strong source of emission in the UV.
The observed flux in November was comparable to that reported by
\cite{slevinsky1999} in the first post-outburst observation in
1978. Therefore it is reasonable to expect that FUV flux from WZ
Sge is dominated by the WD in the 29 September and early November
observations. The main problem with this hypothesis is that the
\cite{cheng1997} analysis of {\it HST} GHRS observations suggest
that the WD in WZ Sge is rotating rapidly ($v \sin{(i)}=1200
\VEL$). If that is correct then the metal lines that are observed
in the {\it FUSE} spectra cannot arise in the WD photosphere,
since they have a velocity width of order 200-400 $\VEL$. As a
result, there are at least three possibilities to explore:

\begin{enumerate}
\item The WD is slowly rotating. If this is correct, then it seems
most likely that the interpretation of the GHRS data was flawed,
presumably due to the limited S/N of the data and to the fact that
there were only a few line features being fitted and that these
were relatively shallow.
\item The WD is rapidly rotating and the narrow lines are due to material
along the line of sight to the WD. If this is correct, then the
narrow lines that are observed in the September and November
observations do not arise from the surface of the WD, but from
material somewhere in the WZ Sge system along the line of sight to
the WD.  In that case, as the system approaches ``complete''
quiescence, the narrow features should disappear and broad
features from the rapidly rotating WD should become apparent.
\item The WD is slowly rotating, and some but not all of the lines
are due to overlying material.
\end{enumerate}

In an attempt distinguish between these choices and to help
extract information about the WD in the event that the lines do
not originate in the WD photosphere, we have created a new grid of
model WD spectra using Ivan Hubeny's TLUSTY and SYNSPEC codes
\citep{hubeny1988,hubeny1995}. The new grid has temperatures from
13,000-80,000 K, $\log{g}$ from 7.5 to 9, and rotation rates
ranging from no rotation to 1500 $\VEL$. It contains models for
pure hydrogen atmospheres and for atmospheres with metal
abundances ranging from 0.01 solar to 10$\times$ solar. To
identify the models which best fit the data, we have conducted
standard $\chi^2$ fits to the data.
%Generally speaking we have performed
%two types of fits to the data, a standard $\chi^2$ fit and one,
%hereafter designated a ``stunted'' $\chi^2$ fit.  The latter is
%designed to provide a qualitatively good fit to the continuum,
%de-emphasizing individual datapoints or small regions that are
%wildly discrepant from the model.  It is a standard $\chi^2$ fit,
%except that we allow no point to contribute more than 25 to the
%overall value of $\chi^2$.
Unless otherwise noted, we have attempted to fit the entire
spectrum, omitting only regions contaminated by airglow or by
obvious emission features, e.g. O VI and C III $\lambda$  977. All
results are quoted in terms of $\chi^2$ per degree of freedom or
$\chi_{\nu}^2$.
%A selection of the better fits are summarized in Table \ref{models}.

We have concentrated our initial spectral fits on the
time-averaged November observation in the hope that contributions
from the disk and any other material in the system would be
minimized then.  The best  $\chi^2$ fit, assuming $\log{g}=8.5$,
is shown in Fig.\ \ref{wzsge4_wd_norm}. The model has a
temperature of 23,200 K, a rotation rate of 190 $\VEL$, and metal
abundances which are 2.2 times solar ratios. Qualitatively, the
model approximates the overall shape of the spectrum fairly well,
although $\chi_{\nu}^2$ is large, 16.9. Assuming the distance is
44 pc and that the WD is fully visible, the normalization implies
a radius for the WD of \EXPU{4.8}{8}{cm}. This radius would
suggest that the mass of the WD is $\sim$1.2 $\MSOL$ (or that the
gravity is actually greater than $\log{g}= 8.5$), but is in any
event consistent with a relatively massive WD. The value of
$\chi_{\nu}^2$ is large mainly because the of discrepancies
associated with the lines in the models. There is a correlation
between gravity and temperature in model fitting of this type
(because both gravity and temperature affect the width and depth
of the Lyman features). When the gravity was set to $\log{g}=8.0$,
the best-fitting temperature becomes 21,500 K, the best rotation
rate remains 190 $\VEL$, and the best metal abundance becomes 1.75
solar. As a result of the fact that the temperature is lower, the
normalization implies a somewhat larger radius \EXPU{5.3}{8}{cm}.
The fits with $\log{g}=8$, are slightly worse in a statistical
sense, $\chi_{\nu}^2$ = 17.6.

There is no particular reason to believe that the metals present
in the  photosphere are solar abundance ratios. One might hope
that a proper choice of abundances could bring the model more into
alignment with the data.  As a result of this, we have
experimented with fitting models in which abundances of individual
elements are varied from the average value of 2.2. The model grids
we created for this purpose had the $\log{g}=8.5$ and a rotational
velocity of 190 $\VEL$. The only element for which there was a
major improvement was for N. Allowing the ion abundance of this
element to vary improved the value of $\chi_{\nu}^2$  to 15.4
(from 16.9). The improvement in $\chi_{\nu}^2$ was due primarily
to improvements in fitting lines at 990, 1084, and 1135 \AA, which
arise from N III, N II, and N I, respectively. The best fit value
for the abundance was 12.5 times solar; the best fit temperature
was essentially unchanged at 23,100 K as expected since the
overall continuum shape is not affected very much by a simple
abundance change.  Fitting models in which C, Si, S, or Fe were
varied did not result in large changes in $\chi_{\nu}^2$, nor in a
qualitative improvement in the fits. Unlike the situation with N,
the best fit abundances were not very far from the value derived
when all of the abundances were varied together.  For example, if
the abundance of C is allowed to vary, the best fit was for an
abundance of 2 times solar.  Inspection of the features which are
sensitive to the C abundance shows that the C III $\lambda$ 1178
is too weak in the best-fit model, but those associated with C II
at 1140 and 1162 \AA\ are too strong.

In addition to problems with specific metal lines, there is also a
discrepancy between the data and the model in the region between
1050 and 1080 \AA\ and also between 990 and 1005 \AA;  these are
regions where quasi-molecular H features are expected.  In our
models, the Lyman$\beta$ quasi-molecular features in the 1050-1080
\AA\ region are included, using data from \cite{allard1999}.
Similar data are not yet available for the Lyman$\gamma$ features
between 990 and 1005 \AA.  Since the theory of the quasi-molecular
H lines has not yet been tested thoroughly in the {\it FUSE}
wavelength range, we have checked and determined that best fitting
values of temperature and metallicity are not altered
significantly if these regions are ignored in fitting the data.

We conclude that if both the continuum and line spectrum of WZ Sge
in the FUV arise in the WD atmosphere, then (a) the WD must be
slowly rotating ($v \sin{(i)} \sim 200 \VEL$), and (b) the
atmospheric abundance of N is enhanced relative to that of other
metals. Both of the results are in conflict with the findings of
\cite{cheng1997}. On the other hand, the abundance pattern we find
would fit with that expected from CNO-processed material, which
has been reported in other WDs in CVs, such as U Gem
\citep{long1999} and VW Hyi \citep{sion2001}.

The extreme alternative to the hypothesis that all of the lines in
the {\it FUSE} spectrum are photospheric is to assume that all of
the lines are created by material along the line of sight to the
WD, but that the WD spectrum is relatively featureless. The
technical problem associated with this hypothesis is that there
are no physically-based models for this material. We do not know
for certain whether it is more likely that this is material
splattered by the disk stream interaction or if it is somehow
associated with the WD itself. Since light from the WD probes a
``pencil beam'', the absorbing material could be located almost
anywhere along the line of sight. Therefore, as an initial
attempt, we have elected to model the absorption in terms of one
or more slabs of overlying material, each of which we assume to be
in LTE. For each slab, there are three parameters, the temperature
$T_s$ of the slab, the column density $N_H$ of the slab, and the
(presumably) turbulent velocity $v_s$ of the slab. To model the
effect of each slab, we have calculated the opacity of LTE
material using TLUSTY/SYNSPEC, convolved the opacities with the
appropriate velocity dependent Gaussian to create a smoothed
opacity, and created grids of optical depths as a function of
column density. This approach is very similar to that developed by
\cite{horne1994} to model 1150-2500 \AA\ {\it HST} spectra of the
high-inclination dwarf nova OY Car in quiescence in terms of a
16,500 K WD and an ``Fe II curtain".  Our grid included
temperatures from 5,000-30,000, densities from $10^9$ to
\POW{13}{cm^{-3}}, turbulent velocities from 50-500 $\VEL$, and
column densities from log($N_H$) of 18 to 22. To restrict the
total number of models, only slabs with solar abundance ratios
were considered.

Attempts to fit the data with a model consisting of a DA WD and a
single slab were unsuccessful. The fits were qualitatively and
quantitatively worse than fits made assuming the lines arise in a
metal-enriched WD photosphere.  The absorption spectrum of a
single slab is simply not rich enough to mimic the lines in the
November {\it FUSE} spectrum of WZ Sge.  (For this reason, it is
unlikely that plausible changes in abundances could substantially
improve the fits.)

On the other hand, as shown in Fig.\ \ref{wzsge4_da_veil}, models
constructed from a DA WD and two slabs of overlying material do
reproduce the data at least as well as the best models based on
white dwarf photospheres.  This particular model was obtained
assuming a velocity broadening of 200 $\VEL$ and a density of
\POW{13}{cm^{-3}} in both slabs. The temperature of the underlying
$\log{g}=8.5$ DA WD is 23,300 K, essentially identical to that
derived for the metal-enriched $\log{g}=8.5$ WD.  The two slabs
have temperatures of 9,000 and 15,000 K and column densities of
\EXPN{1}{20} and \EXPU{3}{20}{cm^{-2}}, respectively. The value of
$\chi_{\nu}^2$ is 10.9, which is considerably better than any of
the metal-enriched photosphere models. Fits obtained from models
assuming lower densities are almost identical to those shown in
Fig.\ \ref{wzsge4_da_veil}; the column densities of the slabs are
about the same, but the temperatures of the slabs are somewhat
lower. The temperature change is a simple reflection of our
assumption of LTE, implying that the ionic abundances are
determined by the Saha equation.

Models constructed from a DA WD and two slabs with lower velocity
dispersion fit the data considerably less well, as might be
expected since the measured line widths are of order 200 $\VEL$.
The sound speed for a plasma with a temperature of 15,000 K is
about 15 $\VEL$, much less that the velocity width of the lines.
This is a challenge that would have to be overcome in the creation
of physical model based on overlying absorption. Most probably one
would need to invoke velocity gradients along the line of sight
since a turbulent plasma with a characteristic velocity of 200
$\VEL$ would be so supersonic that one would expect shocks to
quickly raise the temperature and hence the ionization state of
the plasma beyond what is observed in WZ Sge. An alternative is
that the absorbing material lies just above the WD surface. This
would help to explain the absence of strong orbital modulations of
the lines, but in this picture, which might actually be described
as a WD with an extended atmosphere, one would expect photospheric
lines as well.

%The best stunted $\chi^2$ fit, assuming a velocity of 200 $\VEL$
%and a density of \POW{13} is shown in Fig.\ \ref{wzsge4_da_veil}.
%The temperature of the underlying $\log{g}=8.5$ DA WD is 23,000 K; the
%temperature of the slab is 10,000 K, and the column density is
%\EXPU{3}{20}{cm^{-2}}.  With the notable exception of the region
%between 1120 and 1130 \AA\ and also C III $\lambda$1176, the WD
%plus slab model provides a qualitatively good fit to the data.
%Both of these regions would be affected by lines if the
%temperature of the slab were somewhat higher, but then other
%regions would be fit less well.  Our stunted $\chi^2$ statistic is
%actually better (4.2/dof) than for the normal abundance model.
%(Had a standard $\chi^2$ fit been used, the resulting value of
%$chi^2$ would have been 16.1/dof.)  Fits assuming lower densities
%yield fits that are almost identical to the one shown in Fig.\
%\ref{wzsge4_da_veil}; the column densities are the same though the
%temperature drops to 8,600 and 7,300 at densities of \POW{11} and
%\POW{9} cm$^{-3}$, respectively. This drop in temperature is a
%reflection of the fact that we have assumed LTE, implying that the
%ionic abundances are determined by the Saha equation. One can
%easily imagine that models with more than one temperature would
%improve the overall fit to the data, although in the absence of
%some kind of physical model it is not clear how useful such fits
%would be.

A better fit than for any simple metal-enriched WD photosphere,
even when abundances are allowed to vary, can also be obtained by
assuming that both photospheric lines and absorption by a single
slab contribute to the spectrum. There are a number of
possibilities, depending on whether one assumes that the WD
rotational velocity and the ``so-called'' turbulent velocity of
the veil are identical or not. Restricting ourselves to the
simplest case, the best fitting model with a $\log{g}=8.5$ WD with
solar abundances rotating with $v \sin{(i)} = 200 \VEL$ and a slab
with a turbulent velocity of 200 $\VEL$ and a density of
\POW{13}{cm^{-3}} is shown in Fig.\ \ref{wzsge4_wd_norm_veil}. The
temperature of the WD is 23,100 K, very close to the value in the
models involving a normal abundance WD and the DA WD plus slab.
The temperature of the slab is 15,800 K and the column density
\EXPU{8.5}{19}{cm^{-2}}. The value of $\chi_{\nu}^2$ is 12.4,
compared to 15.4 for our best model based on a metal-enriched WD
photosphere alone.

We have also carried out the same analysis of the 29 September
observation of WZ Sge.  Qualitatively the results are similar. As
shown in Fig.\ \ref{wzsge3_xsol}, the best fit to a WD photosphere
with $\log{g}$ fixed at 8.5, yielded a temperature of 25,200 K,
abundances which were 3.4 times solar, and a rotational velocity
($v \sin{(i)}$) of 250 $\VEL$.  The normalization implies a WD
radius of \EXPU{5.1}{8}{cm}. The value of  $\chi_{\nu}^2$ is 20.2,
slightly greater than for the November observations, but this can
be attributed to the higher S/N in the 29 September observations.
Models with elevated N abundances improve the fits.  In
particular, if all of the heavy element abundances are fixed at
3.4 times solar except N, then the best fitting $\log{g}=8.5$, $v
\sin{(i)} = 250 \VEL$ model has a temperature of 25,000, a N
abundance of 20 times solar, and $\chi_{\nu}^2$ of 18.3. For the
29 September observation, a better fit ($\chi_{\nu}^2$ of 15.8)
can be obtained with the normal abundance WD photosphere and slab.
The temperature of the WD is 25,100 K, the temperature of the slab
is 16,300 K, and the column density is \EXPU{1.0}{20}{cm^{-2}}, if
a $\log{g}=8.5$ WD and a slab density of \POW{13} cm$^{-3}$ are
assumed.

An important point is that for $\log{g}=8.5$ models, the WD
temperature is always close to 25,000 K, regardless of whether a
DA WD, a simple normal abundance WD, or a WD plus slab is assumed.
Thus there is clear evidence of cooling of the WD of approximately
2,000 K degrees in a month. If the WD is less massive the
temperatures of the best fits were lower.  If a $\log{g}=8$ WD is
assumed, then the temperature of the best fit normal abundance WD
plus slab model is 23,500 on 29 September  and 21,000 in early
November. If a $\log{g}=9$ WD is assumed, then the temperature was
26,200 on 29 September and 24,300 in November. The fits themselves
do not distinguish strongly between the various gravities. This is
similar to our analysis of {\it HST} data, which suggests that the
WD cools from 23,000 K on 10 October to 19,000 on 10 November
\citep{sion2002}.

In addition to the WD in the spectra of WZ Sge on 29 September and
in November, there are emission lines due to C III$\lambda$ll77
and to O VI.  A careful inspection of the spectra shows that the
two C III features are centered close to the rest wavelength of C
III, but that the two obvious features associated with O VI are
not. Instead the two obvious features near 1030 \AA\ are actually
centered at 1032 \AA, which is very close to the rest wavelength
of the shorter wavelength line of the O VI doublet. The separation
of the O VI doublet is large enough that a portion of the profile
associated with O VI is actually buried in the continuum near 1038
\AA. Gaussian fits to the profiles indicate widths of 2-3 \AA\
(600-900 $\VEL$) , except for the peak at 1034 \AA, which,
presumably because of contributions from both compontnets of the
doublet, has a FWHM of nearly 4 \AA.  Assuming that the two peaked
structures represent Doppler rotation, the separation of the C II
line, 5 \AA\ implies that $v \sin{(i)}$ for the emission region is
of order $\pm$750 $\VEL$, consistent with a location in the outer
disk.  It is possible, but far from clear that the same material
provides for absorption elsewhere.

\subsection{The rebrightening phase spectrum\label{Rebright}}

As noted previously, the September 7th observation of WZ Sge took
place during a time in which the disk or at least part of it was
in an optical high state.  As a result, it is unclear whether one
should expect the FUV spectrum to be dominated by the WD or by the
disk.  Accordingly, we have attempted to fit the data obtained on
September 7th, both with synthetic WD spectra and with synthetic
disk spectra.  Examples of the fits, in which the fitted regions
were limited to relatively line-free portions of the spectrum, are
shown in Fig.\ \ref{wzsge2_wd_disk}.  While neither model is a
good fit to the data statistically, both do have overall shapes
that plausibly resemble an underlying continuum. It is also
evident, that neither model produces (or could produce) a good fit
to all of the data without the additional components, a veil or
chromosphere.

The relative contributions from the disk and the WD are not
straightforward to determine in the 7 September spectrum. On the
one hand, the spectrum shows absorption lines from the higher
order Lyman series up to at least Lyman$\eta$ (n=7) as expected
from a relatively hot WD; however the flux does not go to zero
near line center as is predicted by the WD model. At 1100 \AA, the
flux from WZ Sge on 7 September was \EXPU{1.2}{-12}{\OIGS}, 2.4
times greater than it would be on 29 September.  Assuming that the
WD was at least as bright on 7 September as on 29 September, then
the WD must contribute at least 40\% of the flux at 1100 \AA\ on 7
September, even though the system was in the rebrightening phase.

To better characterize the WD, under the assumption that the FUV
flux was dominated by the WD on 7 September, we have fit the data
to the $\log{g}=8.5$ grid of synthetic WD spectra, allowing the
temperature, metallicity and rotation rate to vary, and using the
same wavelength regions that were used to fit the 29 September and
November spectra. The best fit has a temperature of 38,600 K, a
rotation rate of 430 $\VEL$, and metal abundances that are
10$\times$ solar, the maximum in our grid. The normalization
implies a WD radius of \EXPU{4.0}{8}{cm}, which is somewhat (10\%)
but not grossly smaller than values found for the radius when the
disk appears optically to be in quiescence. The value of
$\chi_{\nu}^2$ was 26.0 for this fit, not noticeably worse than
similar fits to the 29 September spectra. If the WD dominates,
then the WD photosphere would need to cool by about 15,000 K in
the three weeks between 7 September and 29 September. However, the
metal abundances seem extreme. If the same fits are performed, but
the abundances are constrained to be solar, then the best fit has
somewhat higher temperature of 42,000 K, a lower rotation rate of
270 $\VEL$,  a WD radius of \EXPU{3.3}{8}{cm}, and $\chi_{\nu}^2$
of 36.

The WDs of dwarf novae do respond to dwarf nova outbursts, and a
variety of mechanisms \citep[see, e.g.][]{godon2002} have been
suggested to explain their responses. Temperatures of
39,000-42,000 K are comparable to the temperature of U Gem at the
end of a normal outburst \citep{froning2001} and TT Ari in
quiescence \citep{gaensicke1999}, and somewhat lower than the
value of $\sim$50,000 K for the WD in the nova-like variable and
SW Sex star DW UMa \citep{knigge2000,araujo2003}. So such
temperatures for WZ Sge would not be unprecedented. But it is
unclear whether the normally-lower average temperature of a WD in
these short-period systems can be raised to this high level. (We
do note in passing that if the accretion energy at the WD surface
could heat the surface uniformly, then the accretion rate required
is quite small, coincidentally similar to the value of the disk
accretion rates below.) In the case of U Gem, the WD temperature
declines by about 10,000 K in several months, but here we would
require the WD to cool by 13,000-16,000 K in three weeks. VW Hyi
may be a better analogue. After a superoutburst, the WD cools with
an e-folding timescale of 9.8 days, considerably  more rapidly
than is observed in U Gem \citep{gaensicke1996}. But the peak
temperature for the WD in VW Hyi is 26,400 K and the total
temperature change in 7,090$\pm$430K, considerably less than
required for WZ Sge. On the other hand, the main, or plateau
phase, of the WZ Sge outburst was considerably longer than in a
typical dwarf nova, suggesting that the temperature rise of the WZ
Sge WD might be larger than in a typical dwarf nova system.

If, alternatively, the FUV flux was dominated by the disk, then,
the overall spectral shape suggests $\dot{m}_{disk}$ of about
\EXPU{1.1}{-9}{\MSOL yr^{-1}}, not very different from that during
the 30 July observation. However, the flux on 7 September, was
lower, and the rate required to match the flux at 1100 \AA\ is
\EXPU{2.3}{-10}{\MSOL yr^{-1}} for a system comprised of a 0.8
$\MSOL$ WD and a disk inclined at 75\degr. A model based on summed
stellar atmospheres with this accretion rate would underestimate
the flux at the shortest wavelengths, and if one were confident of
the shape of disk spectra in the FUV would be a fairly strong
argument against a disk origin for the bulk of the FUV emission.
Unfortunately, there is no consensus on the shape of the spectrum
of a steady-state disk in the {\it FUSE} wavelength range. Even if
there were, the disk in WZ Sge is unlikely to be in a steady state
during the rebrightening outbursts.

Some combination of disk and WD emission is also possible.  Model
fitting does not help because there are no simple disk plus WD
models that will match the features that dominate $\chi^2$. At
present, we do not believe one can determine the fraction of
emission due to the disk precisely, since it seems likely that
most of the line features are not created directly above the FUV
photosphere of either the WD or the disk. On the other hand, it is
hard to believe that the WD was intrinsically fainter on 7
September than on 29 September. If that is true, then the WD had
to have been contributing of order 40\% at 1150 \AA\ in early
September.  WD domination in the FUV would not be completely
unprecedented since \cite{froning2001} found that the WD in U Gem
dominated the {\it FUSE} spectrum during the decline from
outburst.

\section{Summary}

We have observed WZ Sge four times during the outburst of 2002. In
that time the spectrum evolved from a disk-dominated to a WD
dominated system.  Our principle conclusions are as follows:

\begin{itemize}
\item A least in the FUV, the outburst luminosity of WZ Sge is low compared to that of
other (mostly longer) period systems for which there are good
distance and inclination estimates.  Based on the strength of the
O VI feature however, the excitation spectrum is high compared to
these systems, suggesting that the boundary layer may be
relatively luminous.
\item The O VI  profile in the outburst spectrum shows clear evidence of outflow, based on the blue
shift of the absorption cores, even in the absence of a classic
P-Cygni profile. In addition to O VI, the outburst spectrum is
rich in absorption lines of moderate ionization state ions of
common elements; this indicates material relatively far from the
WD, suggesting that the disk (or more likely a tenuous layer above
the disk) must be vertically extended.
\item By early September, in the observation occurring during the early rebrightening
phase, the high excitation lines had weakened significantly. It is
not obvious from spectral fits whether disk or WD emission
dominates in the {\it FUSE} wavelength range at this time. It
seems likely that the spectrum arises from a combination of disk
and WD emission, with a minimum contribution of 40\% from the WD.
\item In the 29 September and early November observations, both of which
occurred after the last rebrightening event in the WZ Sge
outburst, the WD dominates the continuum spectrum.  The
combination of parallax distance of 44 pc, the observed UV flux,
and the derived temperature of the WD on 29 September and in
November imply that the WD in WZ Sge is relatively massive, at
least 0.8 $\MSOL$, in agreement with the conclusions of
\cite{skidmore2000} and \cite{steeghs2001}, and possibly larger.
It is interesting that \cite{knigge2002} report weak 6.5 s  as
well 15 s oscillation in {\it HST} spectra of WZ Sge; if the
shorter periodicity is associated with material rotating at
Keplerian velocities, the minimum mass of the WD is 1.03 \MSOL.
For $\log{g}=8.5$, the WD declined by about 2,000 K, from 25,100 K
on 29 September to 22,600 in early November.

\item The lines in the 29 September and early November spectra are narrow with
velocity widths of order 200-400 $\VEL$. If some or all of these
are created in the WD photosphere, as seems natural, then the WD
in WZ Sge is rotating relatively slowly, in contrast to the
conclusion of \cite{cheng1997}. However, it also possible, at
least on the basis of the fits, that all of the lines are created
by material located somewhere along the line of sight to the WD
photosphere.  (The main problem with assuming that material along
the line of sight dominates the absorption spectrum is that there
is no physical model for this material.  In the absence of such a
model, one suspects that one could produce good fits to the
spectra of almost any WD in DN system, thereby bringing into
question a considerable body of research.)
\end{itemize}

This has been a first analysis of the early {\it FUSE}
observations of the outburst of WZ Sge in 2001 July.  A similar
campaign was conducted with {\it HST} \citep{knigge2002,sion2002}.
These along with continued observations of WZ Sge with {\it FUSE}
and with {\it HST} will eventually resolve the question of the
rotation of the WD and with that the degree to which material
along the line of sight to the photosphere contaminated our view
of the WD in the fall of 2002. Ultimately we will obtain a clear
picture of the ultraviolet properties of one of the nearest and
shortest-period dwarf novae.

\acknowledgments{We gratefully acknowledge the efforts of the {\it
FUSE} operation team and the {\it FUSE} Project Scientist George
Sonneborn in making timely observations of the WZ Sge outburst
possible. Our analysis effort was supported by NASA through grants
NAG5-9283 and NAG5-10338. BTG was supported by a PPARC Advanced
Fellowship.

%\appendix
%\pagebreak
%\include{tab1}

\pagebreak

\figcaption[wzsge_obs.ps]{The VSNET measurements of the optical
lightcurve of the 2001 outburst of WZ Sge, as compiled by
\cite{ishioka2002}, with times of the {\it FUSE} (and the {\it
HST}) observations indicated. An expanded view of the observations
that took place when rebrightenings were occurring is shown in the
inset to the figure. \label{wzsge_obs}}

\figcaption[]{ Spectra of WZ Sge as observed with {\it FUSE} in
the summer and fall of 2001. Times past optical maximum and dates
of the observations are indicated. The 30 July spectrum is scaled
so that the OVI line is fully visible. The remaining spectra are
scaled so that the region between 1150 and 1170 \AA\ occupies the
same portion of the plot. The very sharp emission lines seen,
particularly in the November observations, are all geocoronal.
\label{data_fourpanel}
%(Created by macro plotem in sm.four_panel in ~/cv/wzsge/fuse/data)
}

\figcaption[wzsge1_lines.ps]{Expanded view, with lines identified,
of the first {\it FUSE} spectrum of WZ Sge obtained on 30 July
about 6.3 days after optical maximum. \label{wzsge1_lines}}

\figcaption[d1_lif1_O6.ps]{The individual LiF~1 spectra of the O
VI line region as observed on 30 July in time order from bottom to
top. The orbital phase associated with each spectrum is also
shown. The fainter line overlayed on each of the spectra is the
spectrum obtained at phase 0.04, when the flux was least.
\label{d1_lif1_O6} }
%Created with my revised macro sm.lc in directory ~/wzsge/D1new/Comb

\figcaption[wzsge2_lines.ps]{Expanded view with the stronger lines
identified of the{\it FUSE} spectrum of WZ Sge  obtained on 7
September, during the rebrightening phase of the outburst, about
45 days after optical maximum. \label{wzsge2_lines}}

\figcaption[wzsge3_lines.ps]{Expanded view with lines identified
of the {\it FUSE} spectrum obtained on 29 September, several weeks
after the end of the last rebrightening, 67 days after optical
maximum. \label{wzsge3_lines}}

\figcaption[wzsge4_lines.ps]{Expanded view with lines identified
of time-averaged spectrum obtained during the November {\it FUSE}
observation, about three and a half months after the outburst
began. \label{wzsge4_lines}}
%Files for the figures are in figs_apj...but the real work is in cv/wzsge/fuse/data

%\figcaption[wzsge_fuse_linecomp1060_1160.ps]{Expanded view of
%1060-1160 \AA\ spectra of WZ Sge during all four observations
%indicated that the line widths are similar in all four
%observations. \label{linecomp1060_1160}}
%%Files for the figures are in figs_apj...but the real work is in cv/wzsge/fuse/data

\figcaption[data_var4.ps]{LiF~1 spectra of WZ Sge during a 2400 s
interval on 3 November(in grey) and the rest of the November
observation (in black) are shown in the upper panel.  Both spectra
have been binned at 0.5 \AA\ intervals.  A difference spectrum is
shown in the lower panel.\label{data_var4}}

\figcaption[wzsge1_disk.ps]{A comparison of the spectrum of WZ Sge
as observed on 28 July with {\it FUSE} to that expected from a
disk with parameters similar to that of WZ Sge and
$\dot{m}_{disk}$ of \EXPU{8.5}{-10}{\MSOL yr^{-1}} at 44 pc. This
is the $\dot{m}_{disk}$ that matches the flux observed from WZ
Sge. The upper panel shows the model/data comparison.  The lower
panel is the difference between the data and the model.
\label{wzsge1_disk}}

\figcaption[wzsge4_wd_norm.ps]{A comparison of the spectrum of WZ
Sge as observed in November to that expected from a $\log{g}=8.5$
WD with $v\sin{(i)}$ of 190 $\VEL$, a temperature of 23,200 K, and
metal abundances which are 2.2 times solar.  As described in the
text, this was the best-fitting WD model in which $\log{g}$ was
constrained to be 8.5 and the overall abundance was allowed to
vary. The data that were used in the fit are plotted in black; the
data that were omitted, most notably the region around
CIII$\lambda978$, O VI, and various geocoronal lines, are in grey.
\label{wzsge4_wd_norm}}

\figcaption[wzsge4_da_2veil.ps]{A comparison of the spectrum of WZ
Sge in November to that expected from a DA WD and overlying
material. Here, the overlying material was represented in terms of
two slabs, assumed to have densities of \POW{13}{cm^{-3}} and
turbulent velocities of 200 $\VEL$. The resulting fit yielded
temperatures of 9,0000 and 15,000 K and a column densities,
measured in terms $N_H$, of \EXPN{1}{20} and
\EXPU{3}{20}{cm^{-2}}, respectively, for the two slabs.
\label{wzsge4_da_veil}}

\figcaption[wzsge4_wd_norm_veil.ps]{A comparison of the November
WZ Sge spectrum with a model consisting of a normal abundance WD
and a slab.  The WD is assumed to have $\log{g}=8.5$ and the veil
is assumed to have a density of \POW{13}{cm^{-3}}, and turbulent
velocity of 200 $\VEL$.  The best fit has a WD temperature of
23,100 K, a slab temperature of 15,800 K, and a slab column of
\EXPU{8.5}{19}{cm^{-2}}. \label{wzsge4_wd_norm_veil} }

\figcaption[wzsge3_xsol.ps]{A comparison of the spectrum of WZ Sge
as observed at the end of September to that expected from a WD
photosphere. For the fit, $\log{g}$ was fixed at 8.5 and the
temperature and overall metal abundance were allowed to vary.
\label{wzsge3_xsol}}

\figcaption[wzsge2_wd_disk.ps]{A comparison of the spectrum of WZ
Sge as observed on 7 September to that of expected from a WD
(upper panel) and a disk (lower panel). The portions of the
spectrum considered in the fit are shown in black. Thus the fits
are intended to match regions without obvious line absorption.
Other fits for WDs in which fits were made to the entire spectrum
are discussed in the text. Here, a rotation rate for the WD of 200
$\VEL$ and normal abundances were assumed. For the WD, the
temperature was 41,200 K; for the disk, $\dot{m}_{disk}$ was
\EXPU{2.3}{-10}{\MSOL yr^{-1}}. \label{wzsge2_wd_disk}}

\clearpage

% tab1
%\input{tab_obs}
\begin{deluxetable}{lclccccc}
\tablecaption{Observation Summary\label{tab_obs}} \tablewidth{0pt}
\tablecolumns{8} \tablehead{ \colhead{Observation} &
 \colhead{Date} & \colhead{Start}
& \colhead{End} & \colhead{$\delta T_{outburst}$\tablenotemark{a}}
& \colhead{Start Phase\tablenotemark{b}} &
\colhead{End phase} & \colhead{T$_{obs}$} \\
\colhead{} &  \colhead{(2001)} & \colhead{(UT)} & \colhead{(UT)} &
\colhead{(days)}& \colhead{(cycles)} & \colhead{(cycles)} &
\colhead{(s)} } \startdata
1\dotfill &  30 July & 09:36:56 & 11:50:58 & 6.4 &0.47 & 2.11 & 3180  \\
2\dotfill &  7 September & 04:42:43 & 11:53:02 & 45.3 & 0.79 & 6.06 & 8273 \\
3\dotfill &  29 September & 02:50:17 & 08:28:27 & 67.2 &0.49 & 4.64 & 9319 \\
4a\dotfill & 3 November & 05:38:59 & 09:14:51 & 102.3 & 0.98 & 3.63 & 2857 \\
4b\dotfill & 5 November & 14:10:09 & 17:49:56 & 104.7 & 0.53 & 2.22 & 2857 \\
4c\dotfill & 7 November & 07:47:51 & 14:47:34 & 106.5 & 0.12 & 5.27 & 6222 \\
4d\dotfill & 8 November & 08:46:10 & 12:26:30 & 107.4 & 0.48 & 3.18 & 3877 \\
\enddata
\tablenotetext{a}{Time from outburst maximum at midpoint of
observation.} \tablenotetext{b}{The orbital phases of the
observations were calculated using the photometric ephemeris of
Patterson et al.\ 1998.  We applied a phase shift of -0.046 cycles
to correct orbital phase zero from the photometric primary
mid-eclipse to inferior conjunction of the secondary star (Steeghs
et al.\ 2001).}
\end{deluxetable}

\pagestyle{empty}
% fig 1
\begin{figure}
%\plotone{../Figs/wzsge_obs_vsnet.ps}
\plotone{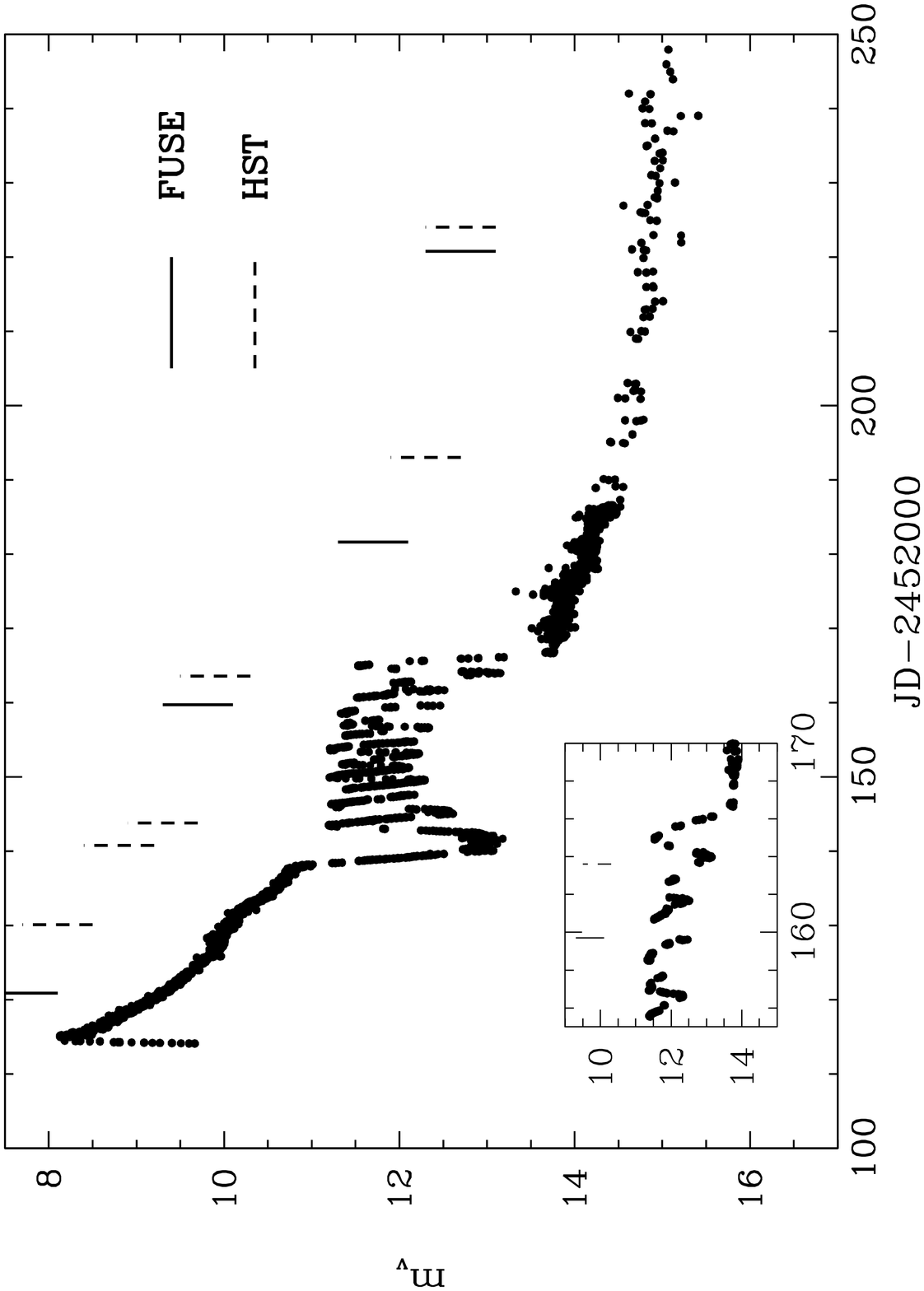}
\end{figure}
\pagestyle{empty}

%fig 2
\begin{figure}
%\plotone{../Figs/data_fourpanel.ps}
\plotone{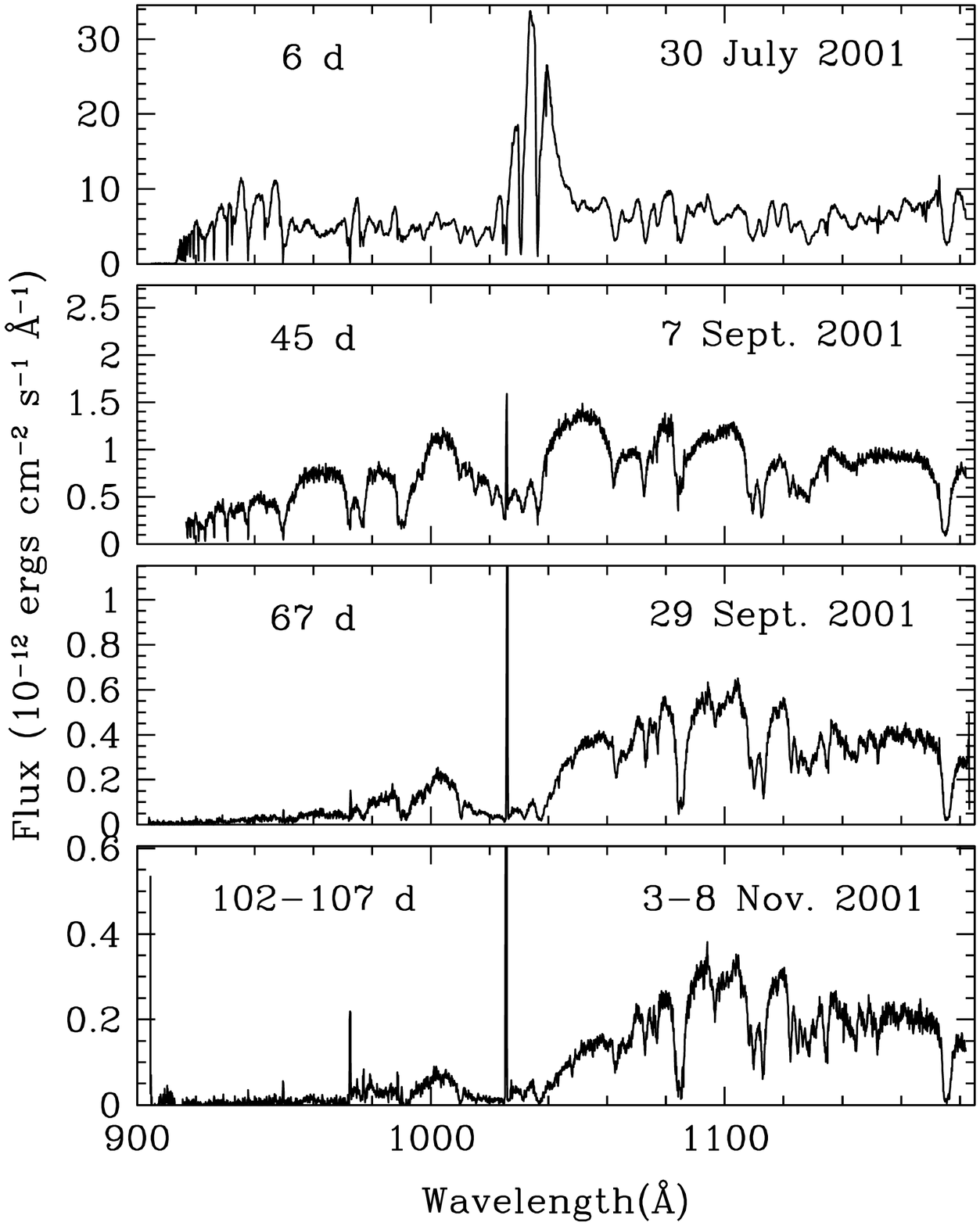}
\end{figure}

%fig 3
\begin{figure}
%\plotone{../Figs/wzsge1_lines.ps}
\plotone{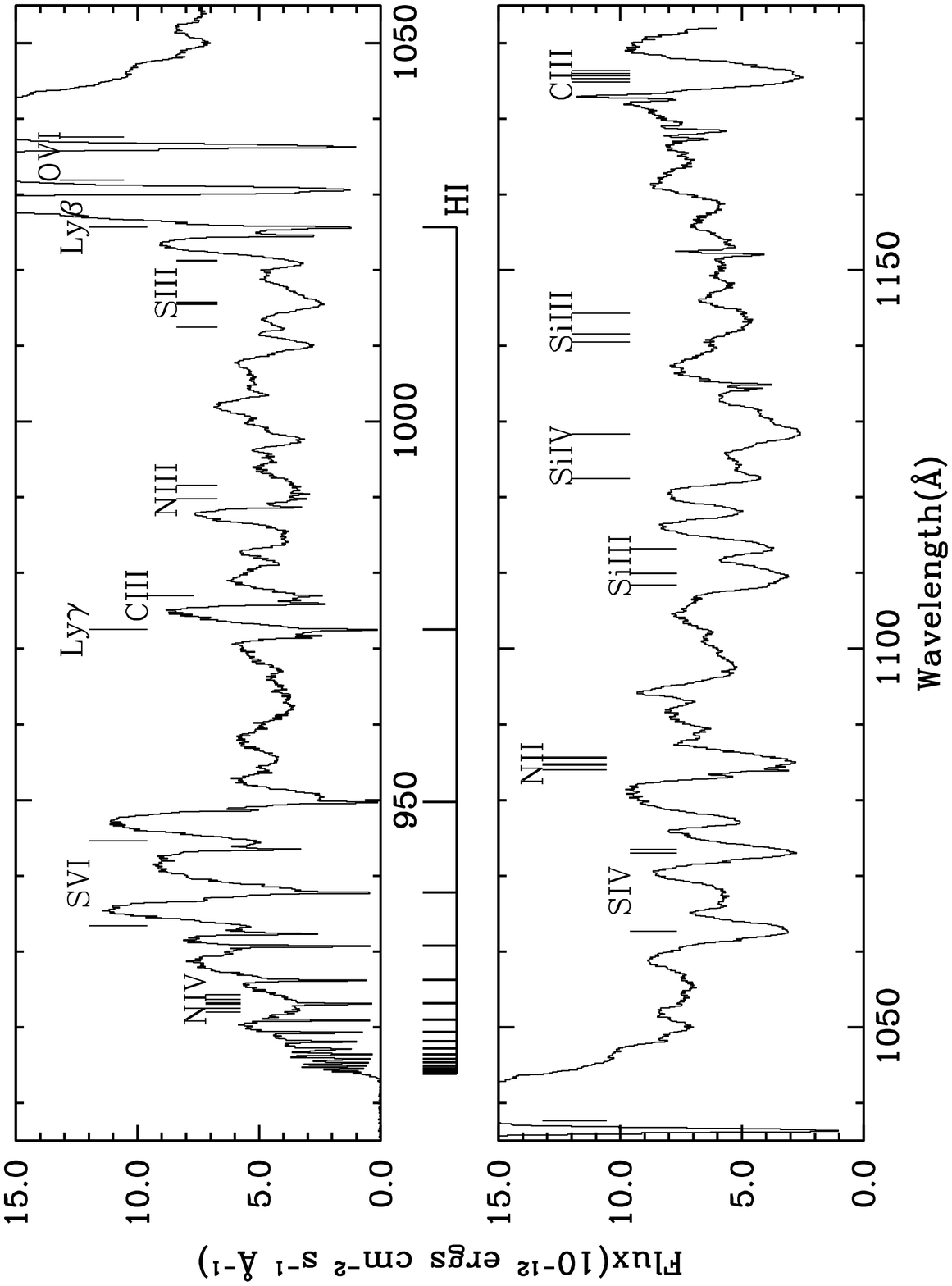}
\end{figure}

%fig4
\begin{figure}
%\plotone{../Figs/d1_lif1_O6.ps}
\plotone{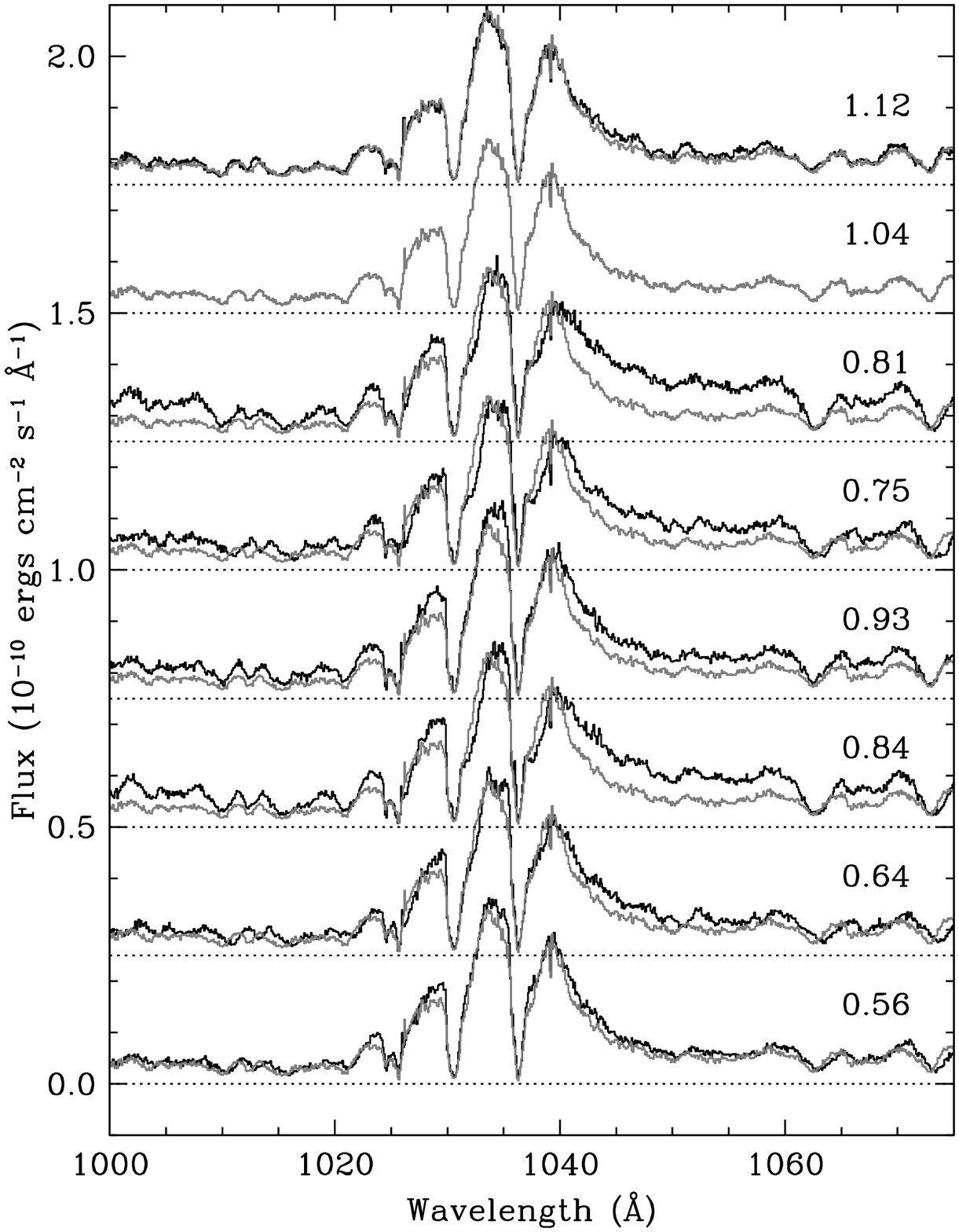}
\end{figure}

%fig5
\begin{figure}
%\plotone{../Figs/wzsge2_lines.ps}
\plotone{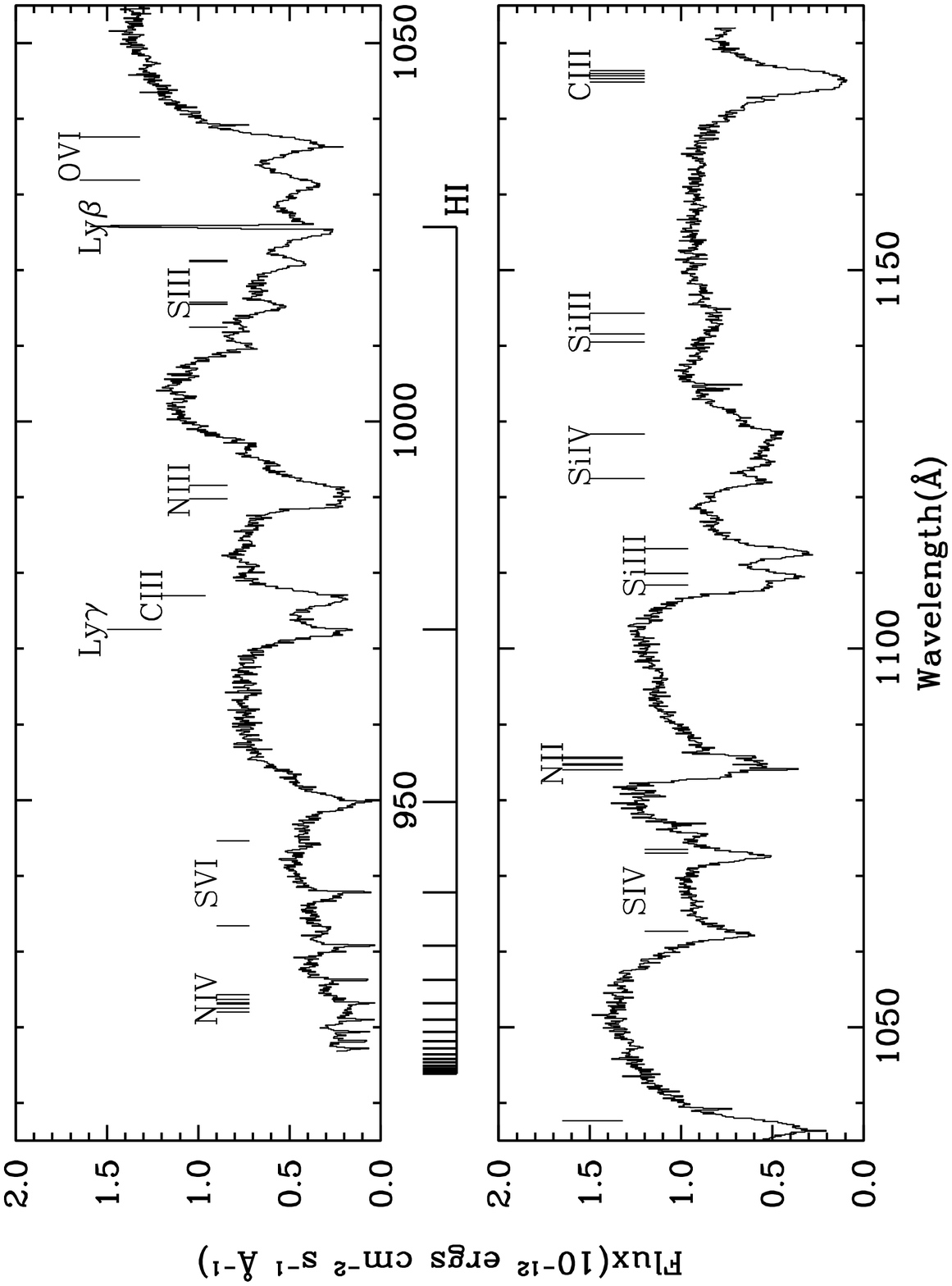}
\end{figure}

%fig6
\begin{figure}
%\plotone{../Figs/wzsge3_lines.ps}
\plotone{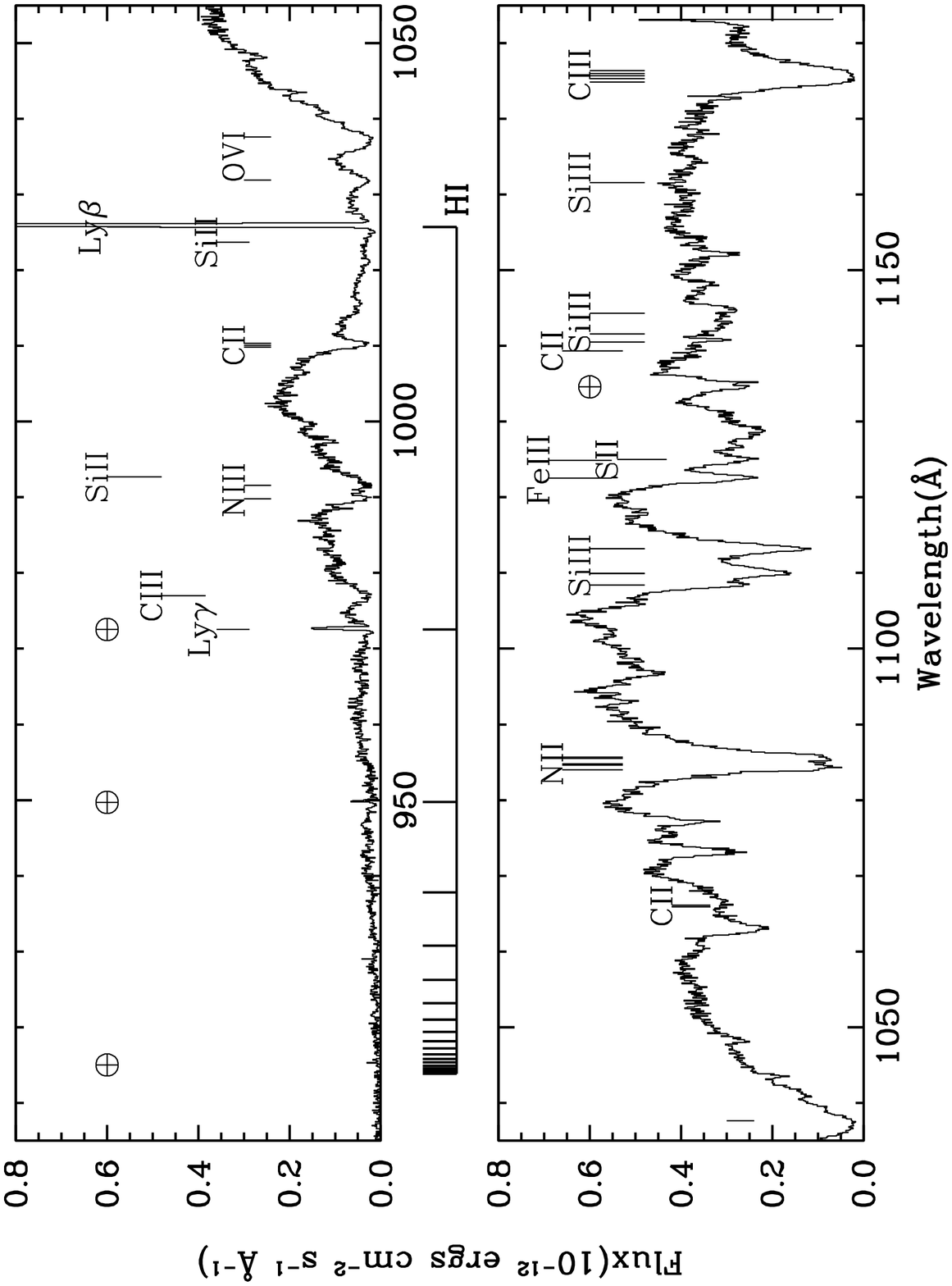}
\end{figure}

%fig7
\begin{figure}
%\plotone{../Figs/wzsge4_lines.ps}
\plotone{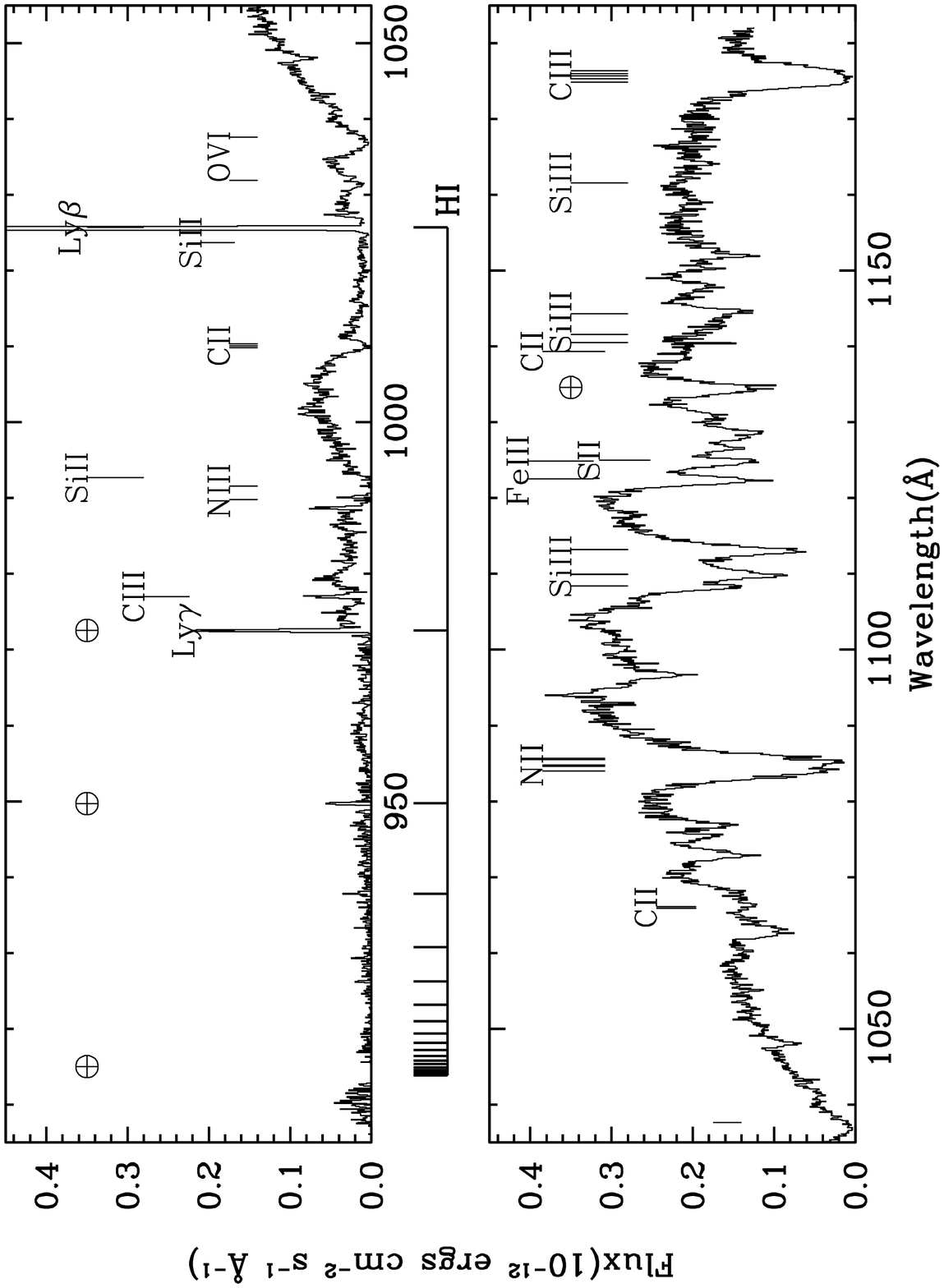}
\end{figure}

%fig8
\begin{figure}
%\plotone{../Figs/data_var4.ps}
\plotone{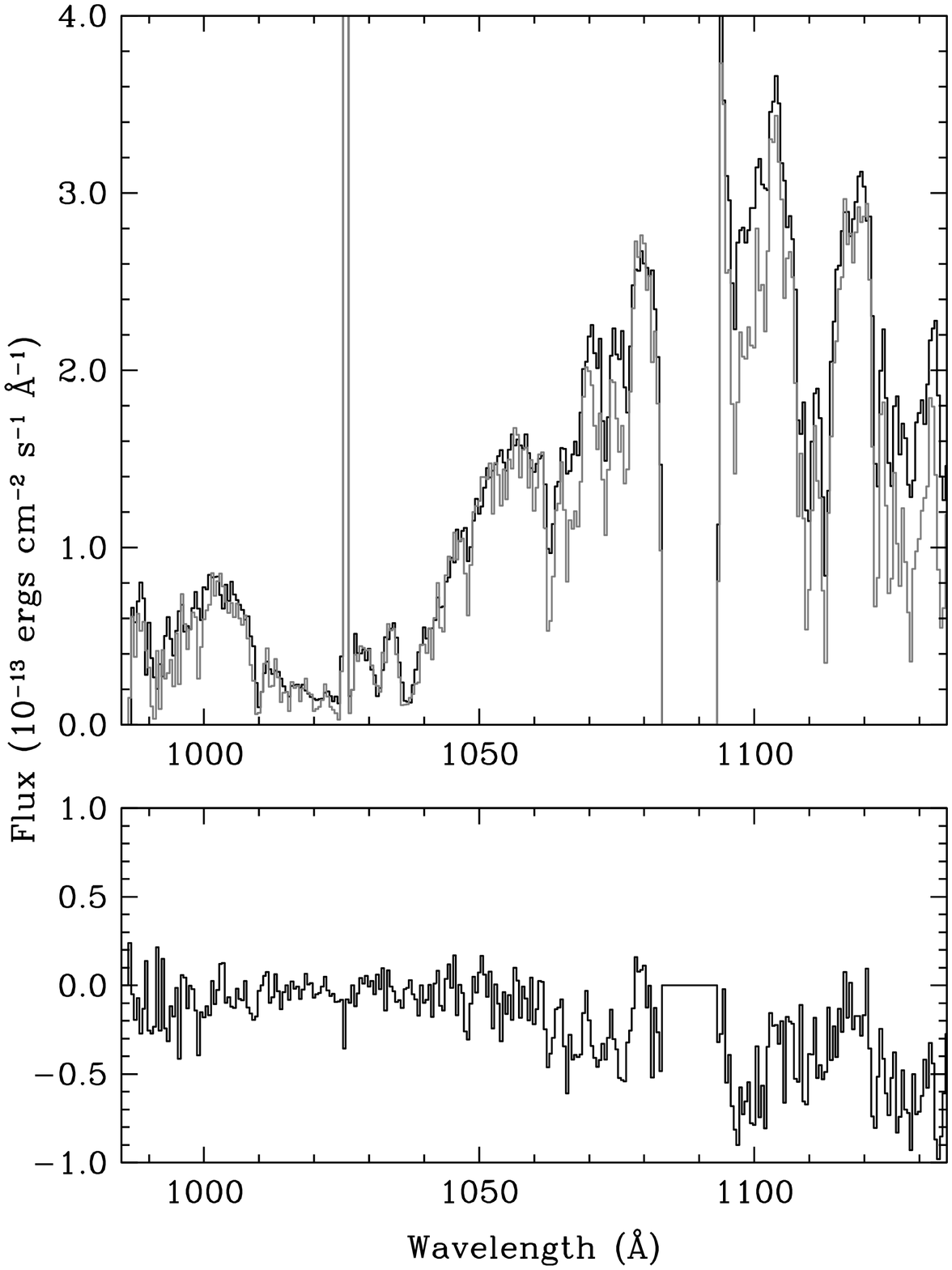}
\end{figure}

%fig9
% Next two figures in ~/cv/wzsge/fuse/figs_apj
\begin{figure}
%\plotone{../Figs/wzsge1_disk.ps}
\plotone{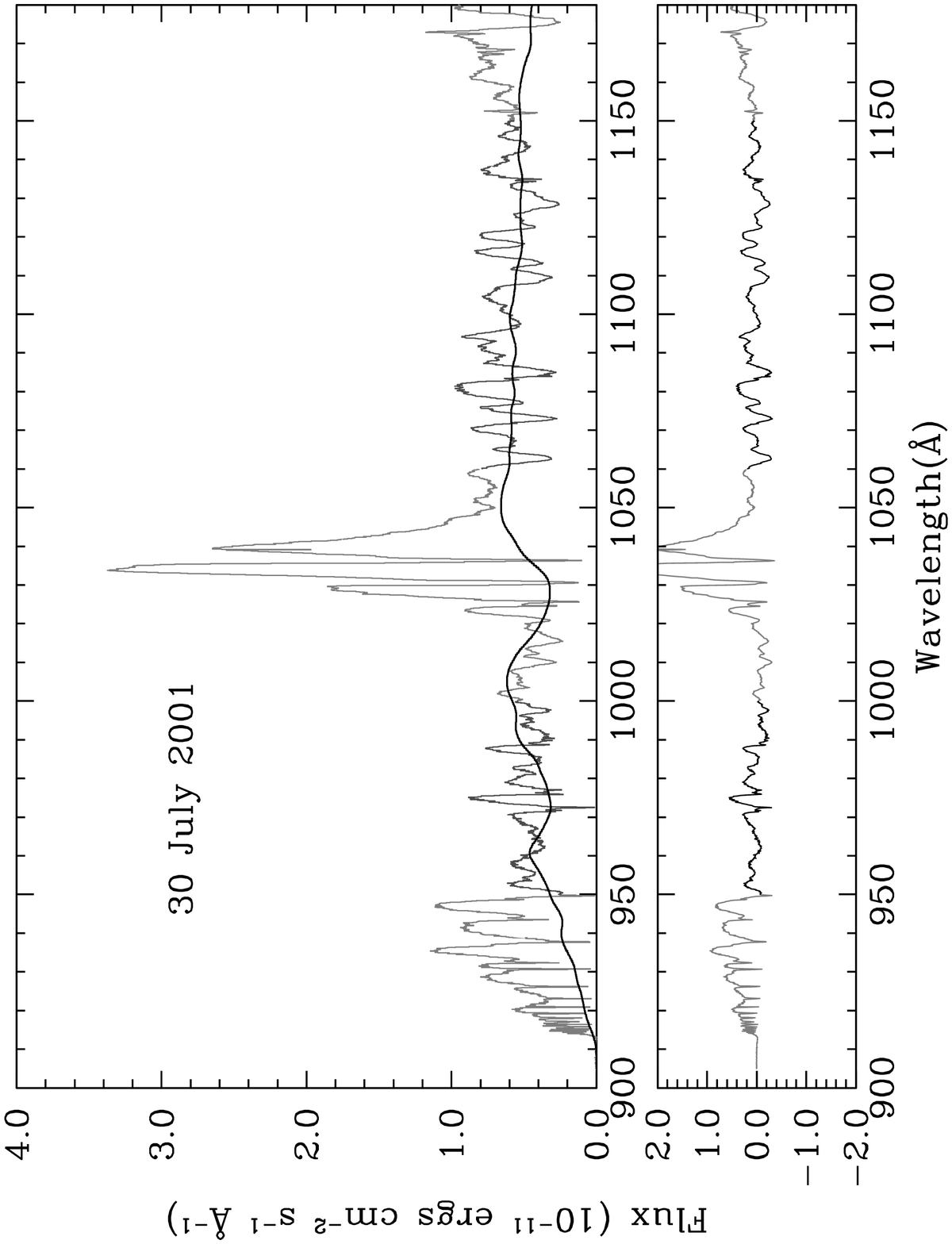}
\end{figure}

%fig10
\begin{figure}
%\plotone{../Figs/wzsge4_xsol.ps}
\plotone{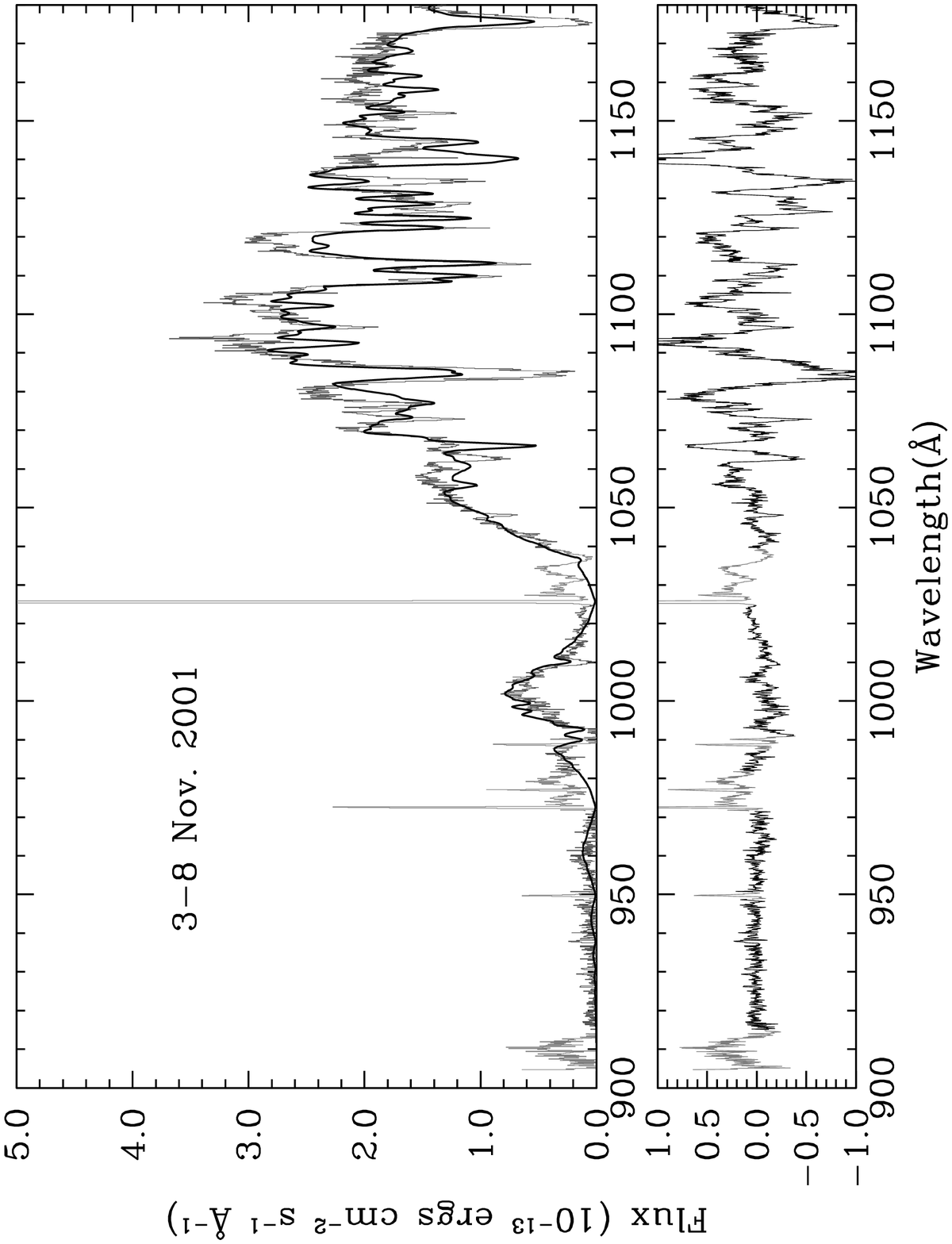}
\end{figure}

%fig11
\begin{figure}
%\plotone{../Figs/wzsge4_da_veil.ps}
\plotone{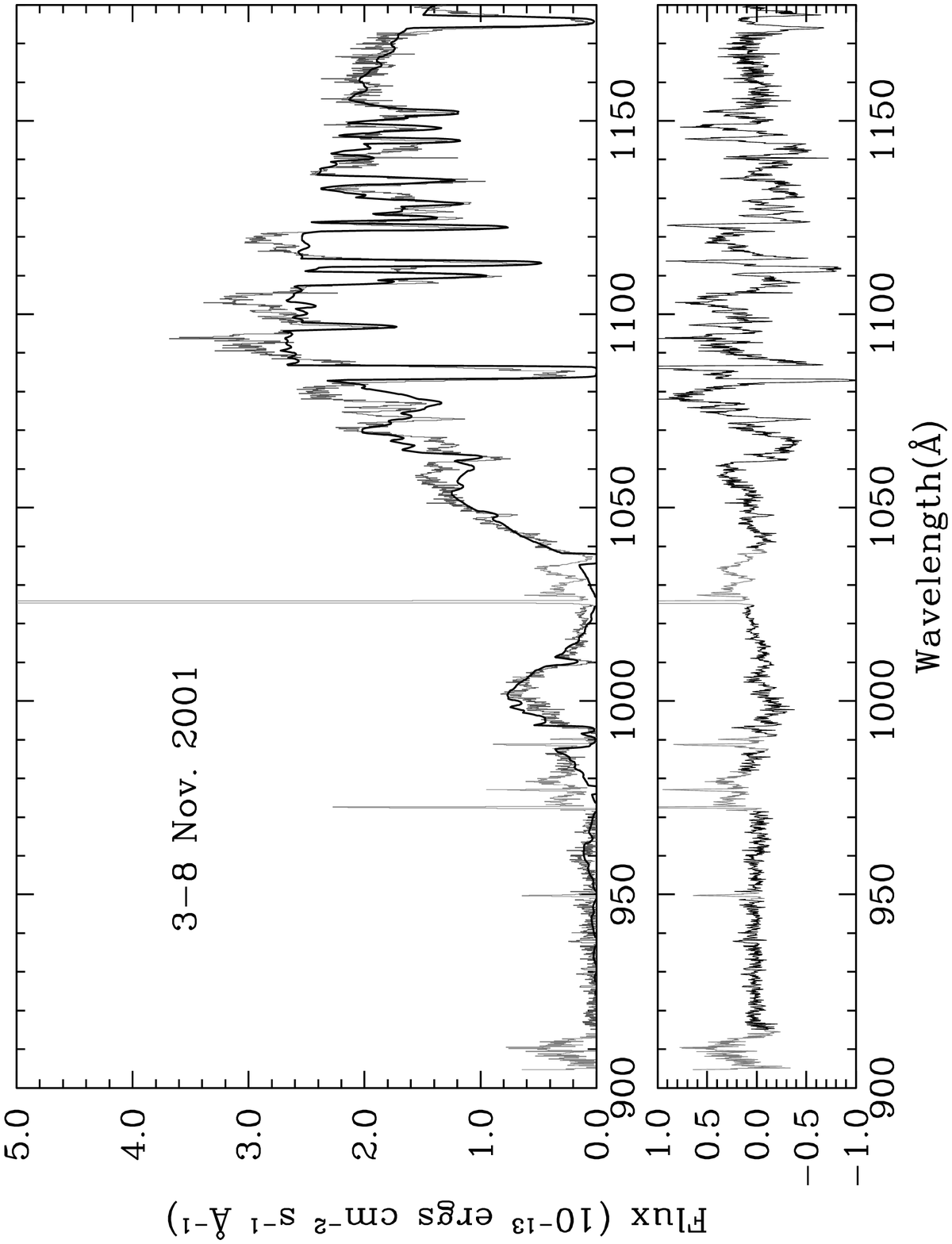}
\end{figure}

%fig12
\begin{figure}
%\plotone{../Figs/wzsge4_norm_veil.ps}
\plotone{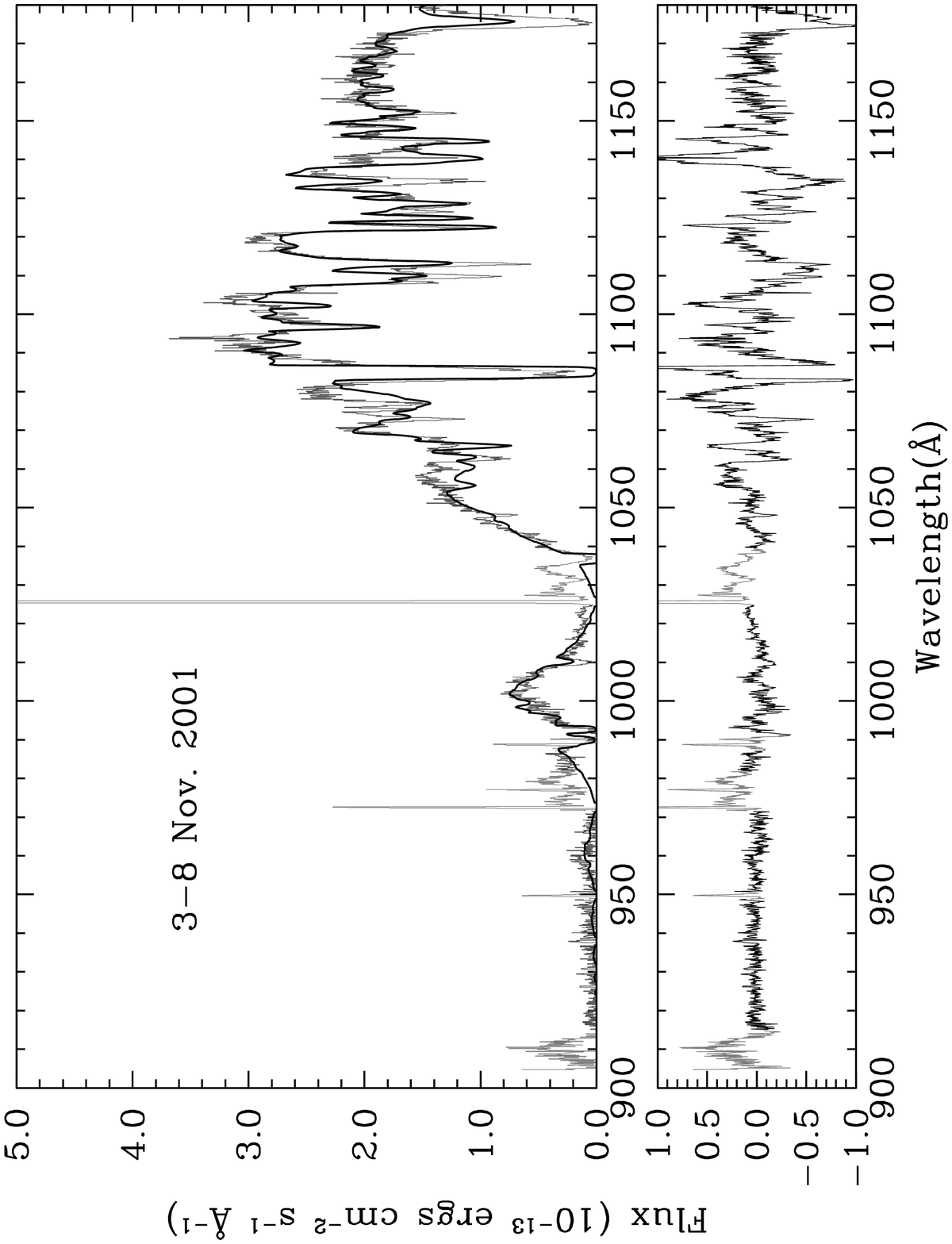}
\end{figure}

%fig13
\begin{figure}
%\plotone{../Figs/wzsge3_xsol.ps}
\plotone{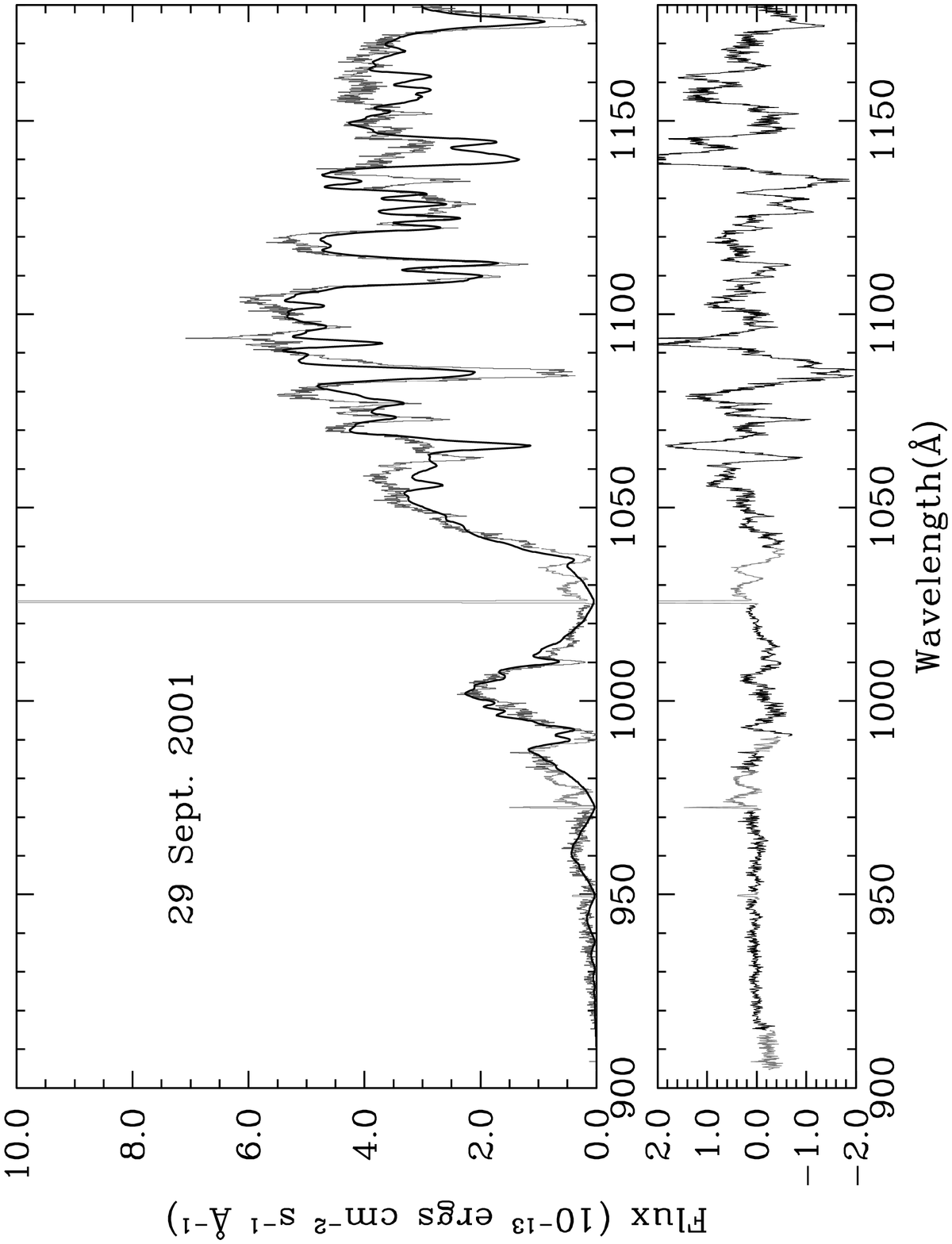}
\end{figure}

%fig14
\begin{figure}
%\plotone{../Figs/wzsge2_wd_disk.ps}
\plotone{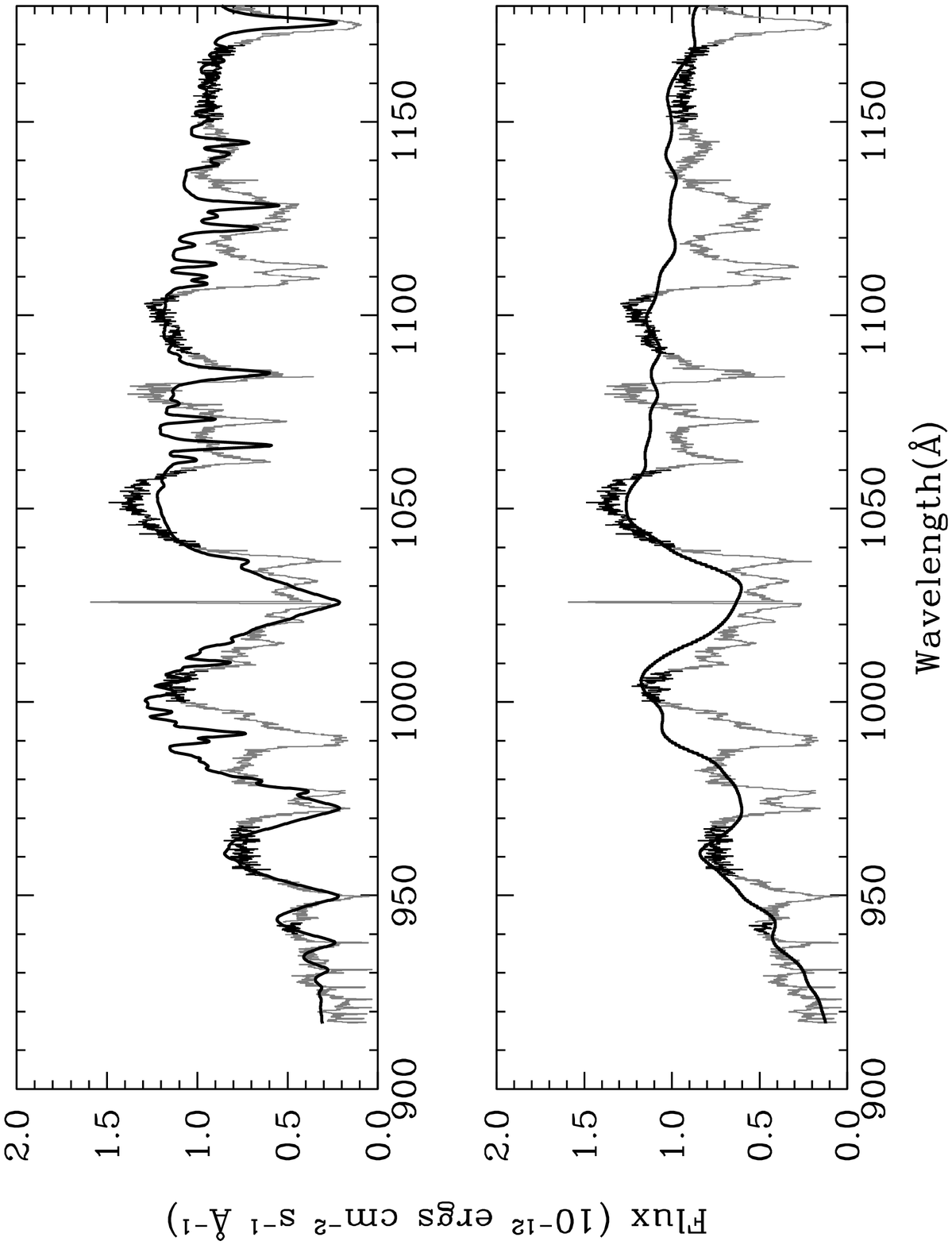}
\end{figure}

\pagebreak
\newpage
\end{document}